\documentclass[]{aa}  
\usepackage{amsmath} 
\usepackage{graphicx} 
\usepackage{epstopdf}
\usepackage{txfonts}
\usepackage{natbib}
\usepackage{mathrsfs} 
\usepackage{fontenc}
\usepackage{bm}  

\newcommand{\dr}[1]{\frac{\partial  #1}{\partial r}}
\newcommand{\dtheta}[1]{\frac{\partial  #1}{\partial \theta}}
\newcommand{\eq}[1]{(\ref{#1})} 

\begin{document}

\title{Gravity darkening in stars with surface differential rotation} 

\author{J. Zorec \inst{1,2}
\and M. Rieutord \inst{3,4}
\and F. Espinosa Lara\inst{5}
\and Y. Fr\'emat \inst{1,2,6}
\and A. Domiciano de Souza \inst{7}
\and F. Royer \inst{8}                        
}
\institute{Sorbonne Universit\'es, UPMC Universit\'e Paris 6 et CNRS, UMR7095 Institut d'Astrophysique de Paris, F-75014 Paris, France
\and 
CNRS, UMR7095, Institut d'Astrophysique de Paris, 98bis Bd. Arago, F-75014 Paris, France; \email{zorec@iap.fr}
\and
Universit\'e de Toulouse; UPS-OMP; IRAP; Toulouse, France
\and
CNRS; IRAP; 14, avenue Edouard Belin, F-31400 Toulouse, France
\and
University of Alcal\'a, E-28871, Alcal\'a de Henares, Spain
\and
Royal Observatory of Belgium, 3 Av. Circulaire, B-1180 Bruxelles, Belgium 
\and
Universit\'e C\^ote d'Azur, Observatoire de la C\^ote d'Azur, CNRS UMR 7293, Lagrange, 28 Avenue Valrose, 06108, Nice Cedex 2, France 
\and
GEPI, Observatoire de Paris, PSL Research University, CNRS UMR 8111, 5 place Jules Janssen, 92190 Meudon
}
 
\offprints{J. Zorec: \email{zorec@iap.fr}}     
\date{Received ..., ; Accepted ...,}
\abstract
{The interpretation of stellar apparent fundamental parameters (viewing-angle dependent) requires that they be treated consistently with the characteristics of their surface rotation law.}
{We aim to develop a model to determine the distribution of the effective temperature and gravity, which explicitly depend on the surface differential rotation law and on the concomitant stellar external geometry.}
{The basic assumptions in this model are: a) the external stellar layers are in radiative equilibrium; b) the emergent bolometric flux is anti-parallel with the effective gravity; c) the angular velocity in the surface obeys relations like {$\Omega(\theta) = \Omega_{\rm o}[1+\alpha\Upsilon(\theta,k)]$} where $\Upsilon(\theta,k)=\cos^k\theta$ or 
$\sin^k\theta$, and where {$(\alpha,k)$} are free parameters.}
{The effective temperature varies with co-latitude $\theta$, with amplitudes that depend on the differential-rotation law through the surface effective gravity and the gravity-darkening function (GDF). Although the derived expressions can be treated numerically, for some low integer values of {$k$}, analytical forms of the ``integral of
characteristic curves", on which the determination of the GDF relies, are obtained. The effects of the quantities {$(\eta,\alpha,k)$} ({$\eta=$} ratio between centrifugal and gravitational accelerations at the equator) on the determination of the {$V\!\sin i$} parameter and on the ``gravity-darkening exponent" are studied. Depending on the
values of {$(\eta,\alpha,k)$} the velocity {$V$} in the derived {$V\!\sin i$} may strongly deviate from the equatorial rotational velocity. It is shown that the von Zeipel's-like gravity-darkening exponent {$\beta_1$} depends on all parameters {$(\eta,\alpha,k)$} and that its value also depends on the viewing-angle $i$. Hence, there no unique interpretation of this exponent determined empirically in terms of {$(i,\alpha$)}.} {We stress that the data on rotating stars should be analyzed by taking into account the rotational effects through the GDF, by assuming {$k=2$} as
a first approximation. Instead of the classic pair {$(\eta,\beta_1)$}, it would be more useful to determine the quantities {($\eta,\alpha,i$)} to characterize stellar rotation.}

\titlerunning{Differential rotation and gravity darkening}
\authorrunning{J. Zorec et al.}
\maketitle

\section{Introduction}\label{intro}

   Rotation induces a geometrical deformation to the stars characterized by a polar flattening and an equatorial stretching. Since the radiation tends to emerge isotropically from the object, the radiation flux becomes a function of the stellar latitude known as the ``gravitational-darkening effect".\par
   According to Poincar\'e-Wavre's theorem, in barotropic systems any of the following statements implies the three others \citep{tass78}: ``1) the angular velocity depends only on the distance $\varpi$ to the rotational axis, $\Omega = \Omega(\varpi)$; 2) the effective gravity is derived from a total gravitation-rotational potential, $\Phi$; 3) the
effective gravity is normal to the isopycnic surfaces; 4) the isobaric and isopycnic surfaces coincide". If the stellar atmospheres of barotropic stars are also in hydrostatic and radiative equilibrium, a relation holds between the bolometric radiation flux $F$ and the effective gravity, $g_{\rm eff}$, known as von Zeipel's theorem \citep{vzpl24}

\begin{eqnarray}
\begin{array}{l}
\displaystyle F = c(\Phi)g^{\beta_1}_{\rm eff},
\end{array}
\label{eq_vz1}
\end{eqnarray}

\noindent where $\Phi$ is the total gravitation-rotational potential describing the stellar surface and $c(\Phi)$ is thus a constant. This relation holds when the radiation flux can be written in the diffusion approximation, which is an asymptotic solution for the radiation field valid at great depths in a semi-infinite atmosphere. In this case, the ``gravity-darkening exponent" takes the value $\beta_1=1.0$.\par
   It was recognized very early on that in rotating objects simultaneous hydrostatic and radiative equilibrium contradict each other \citep{osak66}. However, if radiative equilibrium is maintained forcibly in baroclinic stars, meaning that their rotation laws are non-conservative, $\Omega = \Omega(\varpi,z)$ ($z$ coordinate parallel to the rotation
axis), authors have shown that $\beta_1=\beta_1(\theta)\leq1$, where $\theta$ is the co-latitude \citep{smith74,
kippn77, mae99,love06,clar12}.\par 
   For the particular case of a strict solid-body rotation, \citet{elr11} have shown that Eq.~(\ref{eq_vz1}) can take another formal aspect. Using full two-dimensional stellar models \cite[e.g.][]{RELP16}, they noted that, to a high degree of approximation, the emerging radiation flux-vector $\vec{F}$ emitted by an axially symmetric stellar atmosphere in radiative equilibrium is anti-parallel to the vector of the local effective gravity $\vec{g}_{\rm eff}$. Thus, assuming this parallelism, they write

\begin{eqnarray}   
\begin{array}{l} 
\displaystyle \vec{F} = -\mathscr{F}({\rm r},\theta)\vec{g}_{\rm eff}, 
\end{array} 
\label{eq_vz2}
\end{eqnarray}
 
\noindent where (${\rm r},\theta$) are the spherical coordinates. The function $\mathscr{F}({\rm r},\theta)$ is hereinafter called the gravity-darkening function (GDF). We note that before \citet{elr11},
$\mathscr{F}({\rm r},\theta)$ was simply the constant $c(\Phi)$ in Eq.~(\ref{eq_vz1}). It entered the formulation of the known paradox of von Zeipel, which stipulates the impossibility of having stable barotropic stellar models in radiative equilibrium \citep[e.g.][p. 211]{rxb_66}.\par
  To determine $\mathscr{F}({\rm r},\theta)$ in the case of a rigidly rotating star, \citet{elr11} maintain the condition of radiative equilibrium written in its general form \citep{mih79} and solved the following differential equation 

\begin{eqnarray}
\begin{array}{l}
\displaystyle \nabla\cdot\vec{F} = \vec{g}_{\rm eff}\cdot\nabla\mathscr{F} +\mathscr{F}\nabla\cdot\vec{g}_{\rm eff} 
= 0,
\end{array}
\label{eq_re}
\end{eqnarray} 
   
\noindent and obtained that $\beta_1$ depends on the co-latitude angle $\theta$ and that it is everywhere a decreasing function of the the stellar flattening, $\varepsilon=1-R_{\rm p}/R_{\rm e}$ ($R_{\rm p}$ and $R_{\rm e}$ are polar and equatorial radii, respectively). We note that current interferometric imaging and modelling of rapidly
rotating atmospheres produce inferences of the gravity-darkening exponent in intermediate-mass stars $\beta^{\rm obs}_1\lesssim1$ \citep{monn07,zhao09,che11,monn12,souz14}, although within the current measurement uncertainties, these values are $\beta^{\rm obs}_1\gtrsim\overline{\beta_1}(\varepsilon)$ systematically, where $\overline{\beta_1}(\varepsilon)$ is an average relation between $\beta_1$ and $\varepsilon$ [see Eq.~(\ref{gd_co4}) in 
Sect.\ref{b1theta}].\par
  Apart from the above mentioned radiation transfer reasons, the exponent $\beta_1$ determined observationally can also be dependent on: 1) the line-of-sight angle $i$ of stars, because $\beta_1$ is a function of $\theta$; 2) the differential rotation of the stellar surface, which introduces a stronger dependence of effective temperature with co-latitude than in stars with solid-body rotation. In this respect, \citet{omar13} noted that differential rotation introduces additional contrast on the brightness distribution over the stellar disc, which increases the dependence of $\beta_1$ with the line-of-sight angle.\par 
  We can think of the surface differential rotation as the external imprint of the internal rotation law in baroclinic
stars \citep{rieut07,rieut13}. Statistical inferences concerning intermediate-mass stars suggest that their atmospheres may have differential rotation \citep{zor12}. The rotational increase of the radiative gradient in the envelope of massive rapid rotators significantly enlarges the external convective zones \citep{clem79,mae08}. Like in the Sun, the interaction of rotation with convection may keep driving differential rotation in these regions that in turn can be responsible for the surface differential rotation \citep{zor11,zor17}.\par 
  The aim of the present work is to obtain solutions for the function $\mathscr{F}({\rm r},\theta)$ by imposing the condition of radiative equilibrium in the stellar atmosphere (cf. Eq.~\ref{eq_re}), while the stellar surface rotates differentially. We explore the incidence of this rotation on the observed stellar fundamental parameters, in particular the measured $V\!\sin i$ parameter and the so-called ``gravity-darkening exponent" $\beta_1$.\par

\begin{figure}[t] 
\center\includegraphics[scale=0.72]{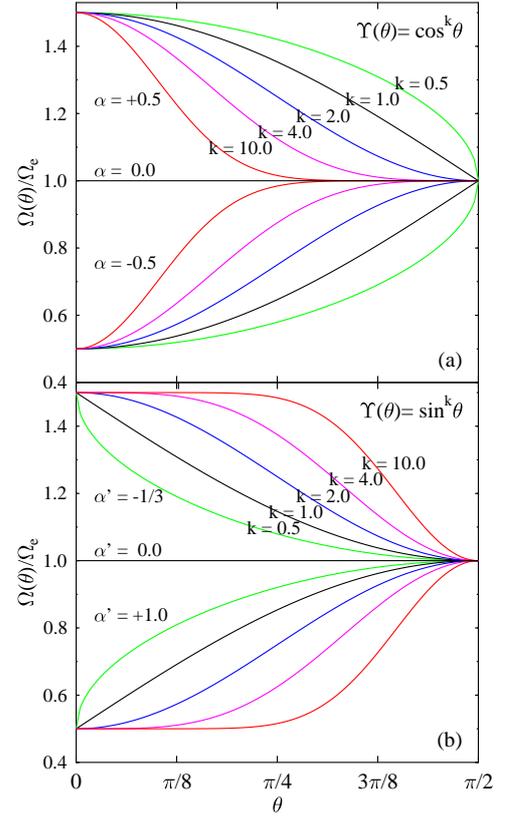}
\caption{\label{cosak} The surface angular velocity $\Omega$ as a function of colatitude $\theta$ as given by Eqs.~(\ref{maund}) and (\ref{maund_l}). (a) $\Upsilon(\theta)=\cos^k\theta$ for $\alpha=+0.5$ and $-0.5$, and several
values of $k$ from 0.5 to 10. (b) $\Upsilon(\theta)=\sin^k\theta$ for $\alpha'=-1/3$ and $\alpha'=+1.0$ $[\alpha'=-\alpha/(1+\alpha]$, and the same values of $k$ as in (a). Colours identify the values of $k$.}
\end{figure}

\begin{figure*}[t]
\center\includegraphics[scale=0.90]{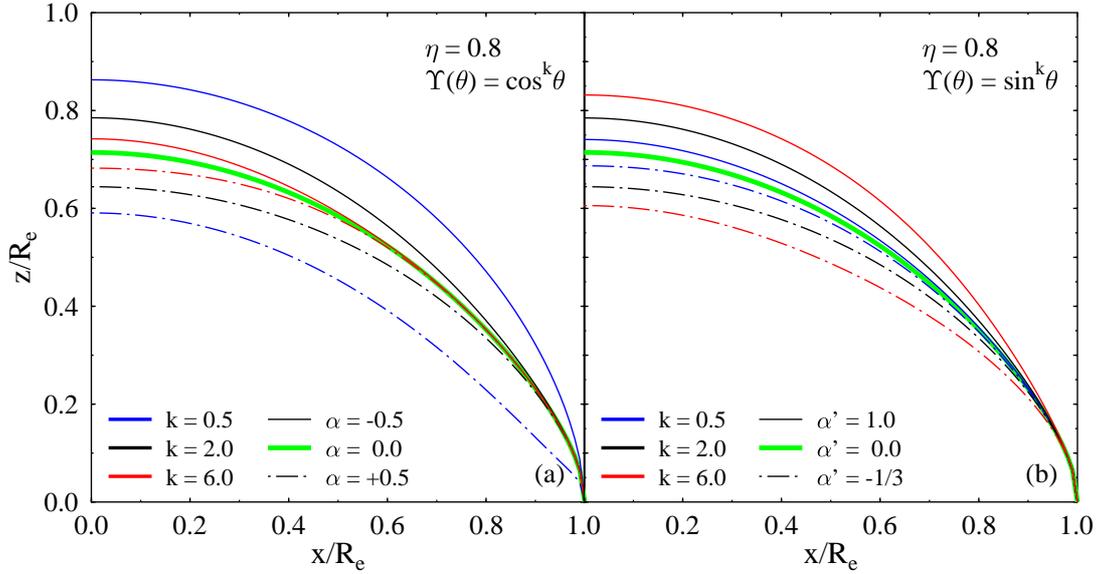}
\caption{\label{fig_geom} Geometrical deformation of stellar surfaces produced by differential rotation  laws given by Egs.~(\ref{maund}) and (\ref{maund_l}). (a) Stellar geometries calculated using $\Upsilon(\theta)=\cos^k\theta$, rotational ratio $\eta=0.8$, differential rotation parameters $\alpha=-0.5$, 0.0 (solid body rotation) and 0.5, and powers $k=0.5$, 2.0 and 6.0. (b) Calculations done with $\Upsilon(\theta)=\sin^k\theta$ and the same values of $\eta$ and powers $k$, but for $\alpha'=-\alpha/(1+\alpha)$. In all cases the colours identify the values of the power $k$, while the line-styles characterize the values of the differential-rotation parameter $\alpha$, and the corresponding 
$\alpha'$.} 
\end{figure*}

\section{The surface differential rotation law}\label{tsdr}

  Having no a priori indications on a general form of the function describing the surface differential rotation law 
$\Omega(\theta)$, but inspired by the simplified Maunder's relation in the solar example \cite[e.g.][p. 819]{maunder_maunder1905}, we propose to use the following relation

\begin{eqnarray}
\begin{array}{l}
\displaystyle \Omega(\theta) = \Omega_{\rm o}[1+\alpha\Upsilon(\theta)],
\end{array}
\label{maund}
\end{eqnarray}

\noindent where $\alpha$ is a free parameter called the ``differential rotation parameter". The case $\alpha=0$ corresponds to rigid rotation. Accordingly, we adopt two ``exploratory" forms for the function $\Upsilon(\theta)$, namely

\begin{eqnarray}
\begin{array}{rcl}
\displaystyle \Upsilon(\theta) & = &
\left\{
\begin{array}{ll} 
\displaystyle \cos^k\theta \\
\displaystyle \sin^k\theta, 
\end{array}
\right.
\end{array}
\label{maund_l} 
\end{eqnarray}

\noindent where $k$ is also a free quantity that takes any real positive value. Figure~\ref{cosak} shows the behaviour of the assumed angular velocity laws given by Eqs.~(\ref{maund}) and (\ref{maund_l}) for $\alpha=+0.5$ ($\alpha'\!=\!-1/3$), $\alpha=0.0$ and $-0.5$ ($\alpha'\!=\!1.0$), and several values of $k$ from 0.5 to 10.0. The higher the value of $k$ the larger the domain sketched as nearly a solid body rotation. This domain occurs near the equator for
$\Upsilon(\theta)=\cos^k\theta$ and near the pole for $\Upsilon(\theta)=\sin^k\theta$. When $\Upsilon(\theta)=\cos^k\theta$ we have $\Omega_{\rm o}=\Omega_{\rm e}$, while for $\Upsilon(\theta)=\sin^k\theta$, $\Omega_{\rm o}=\Omega_{\rm e}(1+\alpha)^{-1}$. In both cases $\Omega_{\rm e}$ is the angular velocity at the equator.\par
   In order to grasp the differences inherent to the use of one of the two $\Upsilon$-functions given in Eq.~\ref{maund_l}, it is necessary to compare the effects produced by rotation laws implying the same ratio 
$\Omega_{\rm p}/\Omega_{\rm e}$. Thus, for a given value of $\alpha$ used with $\Upsilon(\theta)=\cos^k\theta$, we have to take $\alpha'=-\alpha/(1+\alpha)$ associated with $\Upsilon(\theta)=\sin^k\theta$ so that $\Omega(\theta) = \Omega_{\rm o}[1+\alpha'\sin^k\theta]$. Obviously, when $k=2$ both $\Upsilon(\theta)$ represent the same rotation law. Nevertheless, keeping the reciprocity between $\alpha$ and $\alpha'$, the functions $\Upsilon(\theta)=\cos^k\theta$ and $\Upsilon(\theta)=\sin^{k'}\theta$ can produce resembling behaviours of predicted parameters (although not exactly the same), because

\begin{eqnarray} 
\begin{array}{rcl} \displaystyle \sin^{k'}\theta & = & 1-\cos^k\theta \\ 
\displaystyle k'& = & k'(k,\theta),  
\end{array}
\label{maund_s} \end{eqnarray}

\noindent independently of the value of $\alpha$. Taking the average of the two thirds of powers $k'(k,\theta)$ calculated in the middle points of the $\theta-$interval - neglecting the two halves of the remaining third of points situated in the extremes of the interval where in any case the predicted functions fit the right values because of the boundary conditions, it comes, for example that $\Upsilon(\theta)=\cos^k\theta$ with $k=0.5$ will lead to resembling
results as $\Upsilon(\theta)=\sin^{k'}\theta$ when $k'\simeq7.0$, or $\cos^k\theta$ with $k=6.0$ similar results as 
$\sin^{k'}\theta$ where $k'\simeq0.5$ [this effect is illustrated in many figures below, as: Figs.~\ref{fig_tgeff}a and \ref{fig_tgeff}f, or Figs.~\ref{fig_tgeff}c and \ref{fig_tgeff}d; Figs.~\ref{fig_tgeffep}a and \ref{fig_tgeffep}f, or
Figs.~\ref{fig_tgeffep}c and \ref{fig_tgeffep}d; Figs.~\ref{vsini_c}a,d and \ref{vsini_c}b,d, or Figs.~\ref{vsini_s}c,f and \ref{vsini_s}a,c.; Figs.~\ref{fig_b1ie}a and \ref{fig_b1ie}f, or Figs.~\ref{fig_b1ie}c and \ref{fig_b1ie}d].\par 
   In principle we could let $-\infty<\alpha<+\infty$, but we shall limit the values of the differential rotation parameter to $-1\leq\alpha<\infty$ so as to prevent extreme cases where the pole and the equator rotate in opposite senses. Furthermore, according to statistical suggestions based on a study of Be stars \citep{zor17}, in the present work we limit the values of $\alpha$ to the interval $-0.5<\alpha\leq+0.5$. We note that the solar surface differential rotation can be sketched with $\alpha\simeq-0.3$ and $\Upsilon(\theta)=\cos^2\theta$. On the other hand, \citet{rieut13} predicted $\alpha\gtrsim-0.2$ for stars with masses $2\lesssim M/M_{\odot}\lesssim4$. \par  
   Instead of using $\Omega_{\rm e}/\Omega_{\rm c}$ to characterize the rotational velocity at the equator, in this paper we use the ratio $\eta$ between the centrifugal and gravitational acceleration at the equator defined as follows

\begin{eqnarray}
\begin{array}{l}
\displaystyle \eta = \displaystyle  \Omega^2_{\rm e}R^3_{\rm e}/GM = (\Omega_{\rm e}/\Omega_{\rm c})^2\!(R_{\rm e}/R_{\rm c})^3, 
\end{array}
\label{eta}
\end{eqnarray}
 
\noindent where $M$ is the stellar mass, $G$ the graviational constant; $\Omega_{\rm c}$ and $R_{\rm c}$ are the critical angular velocity and the critical radius at the equator, respectively. To make easier any comparison of the expressions derived in the present paper with those in \citet{elr11} where the ratio $V_{\rm e}/V_{\rm K}$ is used as the non-dimensional parameter characterizing the stellar rotation ($V_{\rm e}$ and $V_{\rm K}$ are the actual and Keplerian linear equatorial velocities, respectively), we note that $\eta=(V_{\rm e}/V_{\rm K})^2$, so that it holds $0\leq\eta\leq V_{\rm e}/V_{\rm K}$ $\leq\Omega_{\rm e}/\Omega_{\rm c}\leq1$.\par

\begin{table*}[] \centering \caption[]{\label{applt} Equatorial to polar radii ratio $R_{\rm e}/R_{\rm p}$ as a function of $\eta$, $k$ and $\alpha$, for $\Upsilon=\cos^k\theta$ and $\Upsilon=\sin^k\theta$.}
\begin{tabular}{|r|cccc|cccc||r|cccc|} \hline
 & \multicolumn{4}{c|}{$\Upsilon=\cos^k\theta$} &
 \multicolumn{4}{c|}{$\Upsilon=\sin^k\theta$} & &
\multicolumn{4}{c|}{$\Upsilon=\sin^k\theta$} \\ \hline
\multicolumn{14}{|c|}{$\bm{\eta=0.2}$} \\ \multicolumn{1}{|r}{$\bm{k}$}
&   0.5  &  2.0  & 3.0  & \multicolumn{1}{c}{6.0}&   0.5  &  2.0  & 3.0 & \multicolumn{1}{c}{6.0}  &  \multicolumn{1}{c}{} & 0.5 &  2.0  &  3.0 & 6.0 \\
 $\bm{\alpha}=0.00$ & 1.100 & 1.100 & 1.000 & 1.000 &  1.100 & 1.100 & 1.000 & 1.000 & $\bm{\alpha'}=0.0$ & 1.100 & 1.100 & 1.000 & 1.000 \\
-0.50    & 1.037 & 1.061 & 1.069 & 1.081 &  1.142 & 1.218 & 1.246 & 1.293 & +1.0 & 1.083 & 1.061 & 1.053 & 1.043 \\ -0.25    & 1.065 & 1.078 & 1.083 & 1.090 &  1.113 & 1.134 & 1.142 & 1.154 & +1/3 & 1.091 & 1.078 & 1.074 & 1.068 \\ +0.25    & 1.143 & 1.125 & 1.120 & 1.112 &  1.093 & 1.082 & 1.079 & 1.074 & -1/5 & 1.109 & 1.125 & 1.130 & 1.139 \\ +0.50    & 1.192 & 1.153 & 1.141 & 1.125 &  1.088 &
1.072 & 1.067 & 1.058 & -1/3 & 1.120 & 1.153 & 1.165 & 1.185 \\ 
\hline
\multicolumn{14}{|c|}{$\bm{\eta=0.8}$} \\ 
\multicolumn{1}{|r}{$\bm{k}$} &   0.5  &  2.0  &  3.0  & \multicolumn{1}{c}{6.0}&   0.5  &  2.0  & 3.0 & \multicolumn{1}{c}{6.0} & \multicolumn{1}{c}{} & 0.5 &  2.0  & 3.0 & 6.0 \\ $\bm{\alpha}=0.00$     &  1.400 & 1.400 & 1.400 & 1.400 & 1.400 & 1.400 & 1.400 & 1.400 &  $\bm{\alpha'}=0.00$ & 1.400 & 1.400 & 1.400 & 1.400  \\ 
-0.50 & 1.159 & 1.274 & 1.306 & 1.347 &  1.515 & 1.719 & 1.794 & 1.918 &  +1.0 & 1.350 & 1.274 & 1.247 & 1.202 \\ 
-0.25 & 1.268 & 1.333 & 1.350 & 1.372 &  1.437 & 1.499 & 1.523 & 1.564 & +1/3 & 1.374 & 1.333 & 1.317 & 1.292 \\ 
+0.25 & 1.545 & 1.474 & 1.455 & 1.431 &  1.379 & 1.346 & 1.333 & 1.312 &  -1/5 & 1.427 & 1.474 & 1.491 & 1.521 \\ 
+0.50 & 1.693 & 1.552 & 1.515 & 1.466 &  1.366 & 1.312 & 1.292 & 1.259 &  -1/3 & 1.456 & 1.552 & 1.589 & 1.651 \\
\hline \multicolumn{14}{l}{Note: $\alpha'=-\alpha/(1+\alpha)$; $R_{\rm e}(\alpha=0)/R_{\rm p}(\alpha=0)=1+\eta/2$ given by the Roche gravity-rotation potential that identifies the surface of} \\ 
\multicolumn{14}{l}{rigid rotators. The $\Upsilon=\sin^k\theta$ function is used first with $\alpha$ and then with $\alpha'$.}\\ 
\hline
\end{tabular} \end{table*} 

\section{The stellar geometry and the surface gravity}\label{geom} 

  When the rotation law is conservative, the geometry of the stellar surface can be described with a total gravity-rotation potential. As noted in Sect.~\ref{intro}, the rotation law given in Eq.~\ref{maund} represents the boundary condition in the surface of an internal non-conservative rotational law $\Omega\!=\!\Omega(r,\theta)$. Unlike for barotropic models, in these baroclinic stars it is no longer possible to define a rotational potential. According to \citet{mae09}, the surface in such objects should be the region where an arbitrary displacement $d${\bf s} does not imply any work done by the effective gravity {\bf g}$_{\rm eff}$, i.e. {\bf g}$_{\rm eff}$.$d${\bf s}=0. In the present paper we adopted this approach as it was previously done also in \citet{zor11}.  Figure~\ref{fig_geom} shows the geometries obtained with this method when the rotation law is given by Eqs.~(\ref{maund}) and (\ref{maund_l}), and taking into account several values of $\alpha$ and $k$. The rotation at the equator is characterized by the rotational ratio $\eta=0.8$. In all cases of Fig.~\ref{fig_geom} the colours identify the values of the power $k$, while the line-styles characterize the values of the differential-rotation parameter $\alpha$ (and the corresponding $\alpha'$). The green curves are for $\alpha=0$.\par 
   In Table~\ref{applt} we give the radii ratios $R_{\rm e}/R_{\rm p}$ for $\eta=0.2$ ($\Omega/\Omega_{\rm c}=$ 0.69; $V/V_{\rm c}=$ 0.52) and $\eta=0.8$ ($\Omega/\Omega_{\rm c}=$ 0.99; $V/V_{\rm c}=$ 0.92), for several values of $k$ and $\alpha$. In each series of parameters ($\eta,k,\alpha$) the radii ratios were calculated also for $\Upsilon=\sin^k\theta$ and $\alpha'=-\alpha/(1+\alpha)$.\par
   \citet{zahn10} have shown that the level surfaces of models with shellular rotation laws have the same shape as those for solid-body rotation, including the stellar surface. Moreover, the effect of the internal mass distribution on the surface layer at equatorial critical rotation deviates from a genuine central gravitational field by no more than some 2\%. The same conclusion was also put forward by \citet{zor11} in mo\-dels calculated using Clement's like conservative rotation laws for kinetic-energy parameters higher than $\kappa=K/|W|=0.10$ ($K$ is the kinetic rotational energy stored by the star; $W$ is the gravitational potential energy). Thus, we can safely use the Roche approximation for the surface effective gravity $\vec{g}_{\rm eff}$ and write

\begin{eqnarray}
\begin{array}{rcl}
\displaystyle \vec{g}_{\rm eff} & = & \displaystyle g_{\rm r}\,\hat{\vec{e}}_{\rm r} +g_{\theta}\,\hat{\vec{e}}_{\theta} \\
\displaystyle g_{\rm r} & = & \displaystyle -\frac{GM}{{\rm r}^2} +\Omega^2{\rm r}\sin^2\theta \\
\displaystyle g_{\theta} & = & \displaystyle \Omega^2{\rm r}\sin\theta\cos\theta\,,
\end{array} 
\label{geff}
\end{eqnarray}

\noindent where $G$ is the gravitational constant; $M$ is the stellar mass; $\Omega=\Omega(\theta)$ is the angular velocity given by Eqs.~(\ref{maund}) and (\ref{maund_l}); $\hat{\vec{e}}_{\rm r}$ and $\hat{\vec{e}}_{\theta}$ are the unit vectors associated with the spherical coordinates $({\rm r},\theta)$. In what follows, we use the dimensionless radial coordinate $r={\rm r}/R_{\rm e}$, where $R_{\rm e}$ is the rotationally modified stellar equatorial radius. To simplify notation we write $g_{\rm eff}=|-\vec{g}_{\rm eff}|$ and introduce the dimensionless expression for the effective gravity $\gamma(r,\theta)$ as follows

\begin{eqnarray}
\begin{array}{rcl}
\displaystyle g_{\rm eff} & = & \displaystyle \left(g^2_{\rm r}+g^2_{\theta}\right)^{1/2} = \langle{g}\rangle\gamma(r,\theta) \\
\displaystyle \gamma(r,\theta) & = & \displaystyle \frac{1}{r^2}\left\{1+[(1-\tilde{\eta}r^3)^2-1]\sin^2\theta\right\}^{1/2} \\
\displaystyle \tilde{\eta} & = & \displaystyle \eta\left[\frac{\Omega(\theta)}{\Omega_{\rm e}}\right]^2 \\
\displaystyle \langle{g}\rangle & = &  \displaystyle g_o\left(\frac{R_o}{R_{\rm e}}\right)^2,
\end{array}
\label{geff_2}
\end{eqnarray} 

\noindent where $g_o=GM/R^2_o$ is the surface gravity of the stellar parent non-rotating counterpart having a spherical radius $R_o$ and the same mass $M$ as the rotating object.\par 
   From Eq.~(\ref{geff_2}) it follows that the extreme values of $\gamma(r,\theta)$ are: $\gamma_{\rm e}=\gamma(r,\pi/2)=$ $1-\eta$ and $\gamma_{\rm p}=\gamma(r,0)=$ $(R_{\rm e}/R_{\rm p})^2$, which are functions of $(\eta,\alpha)$ and are dependent on the function $\Upsilon(\theta)$ adopted (see Table~\ref{applt}).\par

\section{The gravity-darkening function}\label{sfgdf}

   The gravity darkening function $\mathscr{F}({\rm r},\theta)$, as defined by Eq.~(\ref{eq_vz2}), is constrained
by the boundary condition obtained when ${\rm r}\to0$, namely 

\begin{eqnarray}
\begin{array}{l}
\displaystyle \lim_{r\to0}\mathscr{F}({\rm r},\theta) = L/4\pi GM,
\end{array}
\label{bc1}
\end{eqnarray} 

\noindent where $L$ is the stellar bolometric luminosity emitted by the stellar core, whose physical properties depend
on the total amount and distribution of the angular momentum inside the star \citep[e.g.][]{sack70,boden71,clem79,
ermu91,urer94,urer95,mema00,deup01,jack05,rieut07,ekst08,mae09,rieut13,fuji15}. In what follows, we use the dimensionless form of $\mathscr{F}$

\begin{eqnarray}
\begin{array}{l}
\displaystyle f({\rm r},\theta) = \mathscr{F}({\rm r},\theta)\frac{4\pi GM}{L}. \\ 
\end{array}
\label{fi} 
\end{eqnarray}

\noindent From Eq.~(\ref{bc1}) we have

\begin{eqnarray}
\begin{array}{l}
\displaystyle \lim_{r\to0}f(r,\theta) = 1.
\end{array}
\label{bc}
\end{eqnarray}  

\begin{figure*}[t]
\center\includegraphics[scale=0.90]{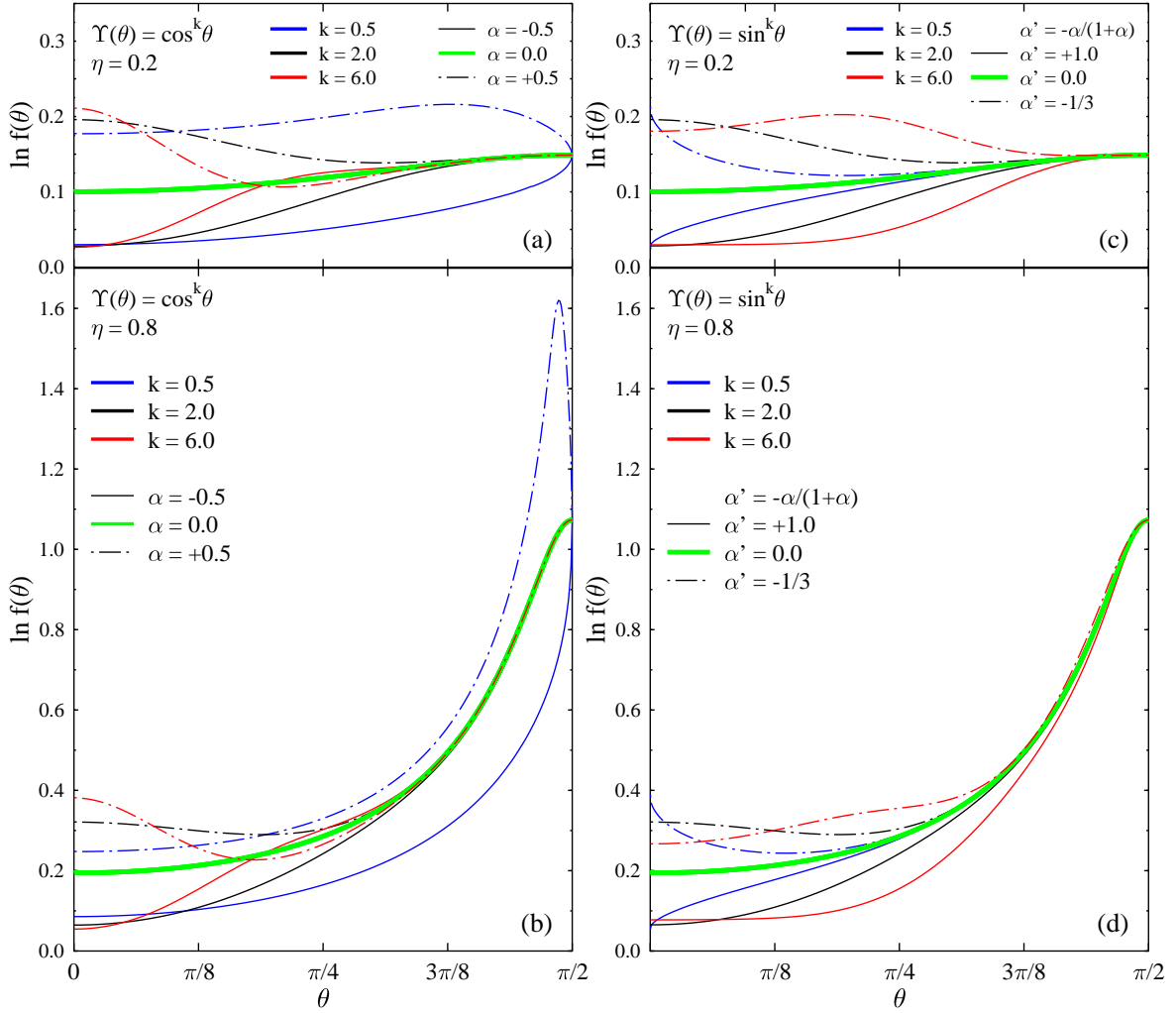}
\caption{\label{fig_frt} Function $\ln f(\theta)$ given by Eq.~(\ref{fi}) against the colatitude $\theta$, calculated 
for $\eta=0.2$ and 0.8, with both functions $\Upsilon(\theta)$ given in Eq.~(\ref{maund_l}), differential-rotation parameters $\alpha=-0.5$, 0.0, and 0.5, and for powers $k=0.5$, 2.0 and 6.0. Blocks (a) and (b) are for 
$\Upsilon(\theta)=\cos^k\theta$. Blocks (c) and (d) are for $\Upsilon(\theta)=\sin^k\theta$. Colours indicate the power $k$ and they are the same for all blocks. The line-styles identify the values of $\alpha$ and they are the same for  
$\alpha'=-\alpha/(1+\alpha)$ where $\alpha$ are the same as in (a) and (b).}
\end{figure*} 
 
\begin{figure*}[]
\center\includegraphics[scale=0.80]{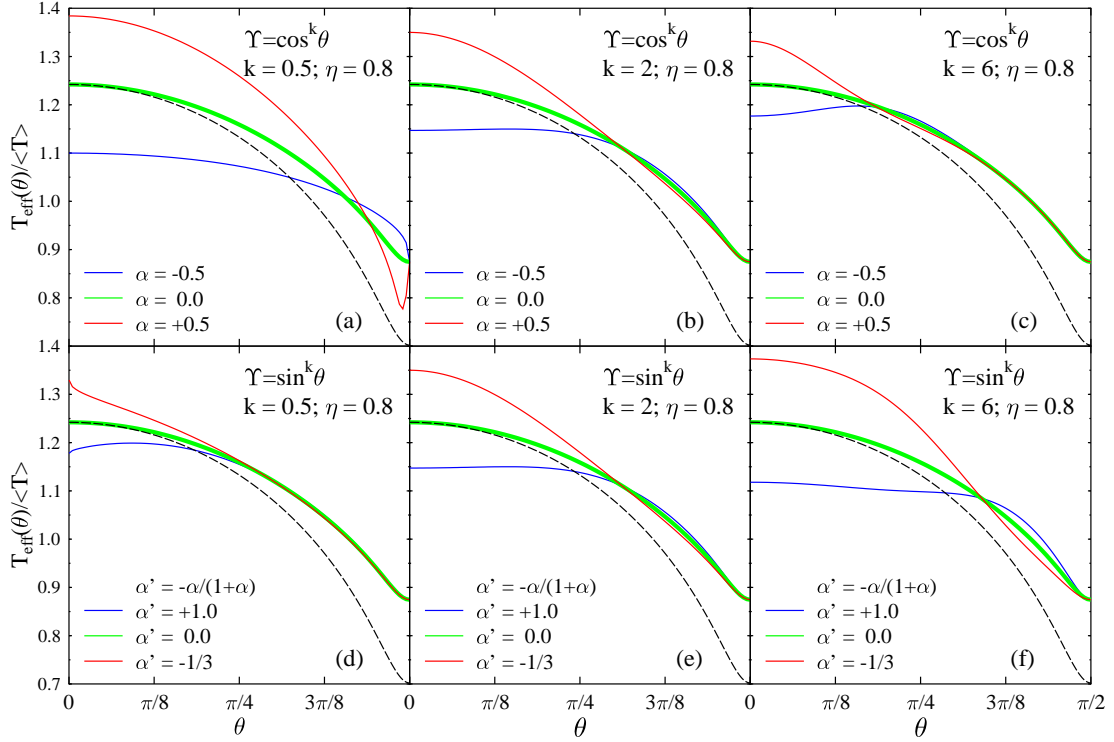}
\caption{\label{fig_tgeff} (a), (b) and (c): Effective temperature $T_{\rm eff}(\theta,\eta,\alpha)/\langle T\rangle$ given by Eq.~(\ref{teff}) as a function of the co-latitude $\theta$ when $\Upsilon(\theta)=\cos^k\theta$, for
$\eta=0.8$, $k=0.5$, 2.0 and 6.0, and $\alpha=-0.5$, 0.0 and +0.5. (d) (e) and (f): Same for $\Upsilon(\theta)=\sin^k\theta$ and the same values of $k$ as before, but for $\alpha'=-\alpha/(1+\alpha)$ where $\alpha$ is the same as in (a) to (c). The black-dashed lines correspond to the classic von Zeipel approximation normalized in $\theta=0$ at the curve corresponding to 
$\alpha=0.0$.}
\end{figure*}  

\begin{figure*}[]
\center\includegraphics[scale=0.80]{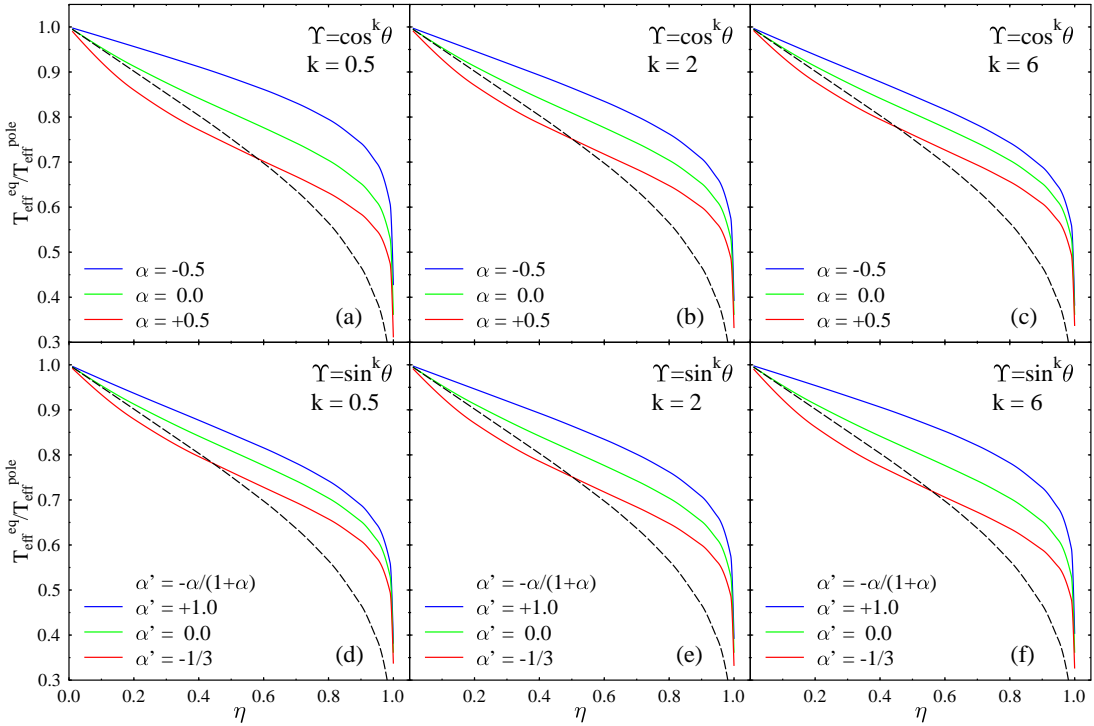}
\caption{\label{fig_tgeffep} (a), (b) and (c): Effective temperature ratios $T^{\rm eq}_{\rm eff}/T^{\rm p}_{\rm eff}$ as a function of $\eta$ and different values $k$ and $\alpha$ when $\Upsilon(\theta)=\cos^k\theta$ (d) (e) and (f): Same for $\Upsilon(\theta)=\sin^k\theta$ and for the same values of $k$ as before, but for $\alpha'=-\alpha/(1+\alpha)$ where $\alpha$ is the same as in (a) to (c). The black-dashed lines correspond to the classic von Zeipel approximation.}
\end{figure*}  

 Using spherical coordinates, the relation in Eq.~(\ref{eq_re}) becomes the following linear partial differential equation:

\begin{eqnarray}
\begin{array}{l}
\displaystyle P\frac{\partial f}{\partial r}+Q\frac{\partial f}{\partial\theta} = 2f, 
\end{array}
\label{pde}
\end{eqnarray} 

\noindent where the notations stand for

\begin{eqnarray}
\begin{array}{rcl}
\displaystyle P & = & \displaystyle -\frac{g_r}{\Omega\left[\Omega+
(\partial\Omega/\partial\theta)\sin\theta\cos\theta\right]} \\
\displaystyle Q & = & \displaystyle -\frac{g_{\theta}}{r\Omega\left[\Omega+
(\partial\Omega/\partial\theta)\sin\theta\cos\theta\right]}. 
\label{rdfe} 
\end{array}
\end{eqnarray} 

   The integration of Eq.~(\ref{pde}) can be reduced to the integration of Lagrange's system  of ordinary linear differential equations, called also ``method of characteristics". Let us follow \cite{rieut15}. We first eliminate the RHS of Eq.~(\ref{pde}) by writing the equation for $\ln f$ and setting :
   
\begin{eqnarray}
\begin{array}{l}
\displaystyle \ln f = \ln G+A(\theta)\;.
\label{new_eq1}
\end{array}
\end{eqnarray}

   After some calculations we find that

\begin{eqnarray}
\begin{array}{l}
\displaystyle A(\theta) = -\ln\left[\Omega^2(\theta)\tan^2\theta\right].
\label{new_eq2}
\end{array}
\end{eqnarray}

   Now, Eq.~(\ref{pde}) may be rewritten

\begin{eqnarray}
\begin{array}{l}
\displaystyle P\dr{G}+Q\dtheta{G}=0.
\label{new_eq3}
\end{array}
\end{eqnarray}

  The characteristic lines of $G$ in the $(r,\theta)$-plane, are the lines where $G$ is constant. On these lines 
$\partial_rGdr+\partial_\theta Gd\theta=0$, so that on these lines we have

\begin{eqnarray}
\begin{array}{l}
\displaystyle Qdr-Pd\theta = 0,
\label{new_eq4}
\end{array}
\end{eqnarray}

\noindent which is the equation of characteristic lines. After some simplifications it reads

\begin{eqnarray}
\begin{array}{l}
\displaystyle -(g_\theta/r) dr+g_rd\theta = 0.
\label{charlines}
\end{array}
\end{eqnarray}

   To solve \eq{charlines}, we need to transform it into an exact differential. We can indeed multiply the equation by any function of $(r,\theta)$. It turns out that if we multiply \eq{charlines} by 

\begin{eqnarray}
\begin{array}{l}
\displaystyle M(r,\theta)=N(\theta)\frac{\Omega^2(\theta)r^2}{GM}\; ,
\label{new_eq5}
\end{array}
\end{eqnarray}

\noindent it can be integrated. Indeed, demanding that  

\begin{eqnarray}
\begin{array}{l}
\displaystyle \dtheta{(-Mg_\theta/r)} = \dr{(Mg_r)},
\label{new_eq6}
\end{array}
\end{eqnarray}

\noindent leads to the following expression for $N(\theta)$:

\begin{eqnarray}
\begin{array}{l}
\displaystyle N(\theta) = A \frac{\cos^2\theta}{\Omega^4(\theta)\sin\theta},
\label{new_eq7}
\end{array}
\end{eqnarray}

\noindent where $A$ is an arbitrary constant. \par
    Thus $G$ is constant on the lines $\tau(r,\theta)=$ Cst, if $\tau$ verifies:

\begin{eqnarray}
\begin{array}{l}
\displaystyle \frac{\partial\tau}{\partial{r}} = -\frac{Mg_\theta}{r} =
\frac{r^2\Omega^2_{\rm o}}{GM}\cos^3\theta \\
\displaystyle \frac{\partial\tau}{\partial\theta} = M g_r=
\frac{\cos\theta\cot\theta}{\left[1+\alpha\Upsilon(\theta)\right]^2}-\frac{r^3\Omega^2_{\rm
o}}{GM}\sin\theta\cos^2\theta, 
\end{array}
\label{dtautt}
\end{eqnarray}

\noindent where we chose $A=-\Omega^2_{\rm o}$. Since the RHS of the previous equations have been chosen such that $d\tau$ is an exact differential, the integration is straightforward and gives 

\begin{eqnarray}
\begin{array}{l}
\displaystyle \tau(r,\theta) = \frac{1}{3}\eta\left(\frac{\Omega_{\rm o}}{\Omega_{\rm e}}\right)^2 r^3\cos^3\theta + \mathscr{T}(\theta)\;.
\label{tau}
\end{array}
\end{eqnarray}

   The function $\mathscr{T}(\theta)$, here referred to as the {\it integral of characteristic curves}, is finally given by

\begin{eqnarray}
\begin{array}{l}
\displaystyle \mathscr{T}(\theta) = \int_{\pi/2}^{\theta}
\frac{\cos\phi\cot\phi}{\left[1+\alpha\Upsilon(\phi)\right]^2}d\,\phi
\end{array}
\label{intau}
\end{eqnarray} 

\noindent that can be calculated numerically for any real positive value of $k$. For some low natural number $k$, the integral in Eq.~(\ref{intau}) admits analytical expressions that are given in appendix A.\par 
   In this work we used only numerical estimates of $\mathscr{T}(\theta)$. The numerical estimates of the integral in
Eq.~(\ref{intau}) were obtained proceeding to the change of variable $u=\ln\tan(\phi/2)$ and using the 16-point Gauss-Legendre quadrature rule. The function $\mathscr{T}(\theta)$ thus obtained was also calculated for all values of $k$ for which we could obtain the analytical expressions given in Appendix~\ref{tft}. We thus noted that the errors of the numerical
estimates of Eq.~(\ref{intau}) are $\delta\mathscr{T}\lesssim10^{-6}$ at $\theta\sim10^{-5}$ rad ($\theta\sim0^\circ.001$), and they become $\delta\mathscr{T}\lesssim10^{-12}$ as soon as $\theta\gtrsim0.07$ rad
($\theta\gtrsim8^\circ$). The precision of the numerical estimates
increases even more as $\theta\to\pi/2$.\par
  The limit expression of $\mathscr{T}(\theta)$ for $\alpha=0$ comes  directly from Eq.~(\ref{intau}), which is also obeyed by all analytical expressions given in Appendix~\ref{tft}:

\begin{eqnarray}
\begin{array}{l}
\displaystyle \lim_{\alpha\to0} \mathscr{T}(\theta) = \ln\tan\frac{\theta}{2}+\cos\theta,
\end{array}
\label{limtau} 
\end{eqnarray}
  
\noindent which is the form of $\mathscr{T}(\theta)$ obtained by \citet{elr11} [see their Eq.~(20)].\par
   Because $\tau\equiv\tau(r,\theta)$, where $r$ and $\theta$ are independent variables, we can use instead the pairs $(r,\tau)$ or $(\theta,\tau)$ as independent variables to obtain two simplified forms of Eq.~(\ref{pde}) valid over the $\tau-$characteristic curves

\begin{eqnarray}
\begin{array}{l}
\displaystyle Q\left(\frac{\partial f}{\partial\theta}\right)_{\tau} = 2f \ \ \ {\rm and} \ \ \ P\left(\frac{\partial f}{\partial r}\right)_{\tau} = 2f. 
\end{array}
\label{tpde} 
\end{eqnarray}

  The first differential equation in Eq.~(\ref{tpde}) can be easily integrated to obtain

\begin{eqnarray} 
\begin{array}{l}
\displaystyle f(r,\theta) = \frac{\Psi(\tau)}{\tan^2\theta[1+\alpha\Upsilon(\theta)]^2},
\end{array} 
\label{solf_t} 
\end{eqnarray}

\noindent where the factor $\Psi(\tau)$ is an integration constant that carries the dependence of $f$ with $r$ through a transcendental function $\vartheta(r,\theta)$ defined, using Eq.~(\ref{tau}), as

\begin{eqnarray}
\begin{array}{l}
\displaystyle \mathscr{T}(\vartheta) = \frac{1}{3}\eta\left(\frac{\Omega_{\rm o}}{\Omega_{\rm e}}\right)^2 r^3\cos^3\theta + \mathscr{T}(\theta),
\end{array} 
\label{ctau} 
\end{eqnarray}

\noindent which assumes that there is a unique relation between $\vartheta$ and $\tau$. From Eq.~(\ref{ctau}) it appears that 
$\vartheta\to\theta$ as $r\to0$. Therefore, noting that $\tau(r,\theta)$ is constant over a characteristic curve, the value of $\Psi(\tau)$ can be specified at any point $(r,\theta)$ on the characteristic, in particular at $r=0$. Thus, making use of the boundary condition in Eq.~(\ref{bc}) we obtain

\begin{eqnarray}
\begin{array}{l}
\displaystyle \Psi(\tau) = \tan^2\vartheta\,\left[1+\alpha\Upsilon(\vartheta)\right]^2.
\end{array} 
\label{psi_t} 
\end{eqnarray}
Introduced into Eq.~(\ref{solf_t}), it leads to

\begin{eqnarray}
\begin{array}{l}
\displaystyle f(r,\theta) = \left(\frac{\tan\vartheta}{\tan\theta}\right)^2\left[\frac{1+\alpha\Upsilon(\vartheta)}{1+\alpha\Upsilon(\theta)}\right]^2.
\end{array}
\label{f_th} 
\end{eqnarray} 

    Eq.~(\ref{f_th}) is the generalized version of that obtained by \citet{elr11}, which is valid for $\alpha=0$ [their Eq.~(26)].\par
    Equations~(\ref{tau}) and (\ref{ctau}), together with Eq.~(\ref{intau}), or the forms for $\mathscr{T}(\theta)$ given in Appendix~\ref{tft}, define $\vartheta(r,\theta)$ for every chosen point $(r,\theta)$, in particular for the stellar surface represented by the function $r=r(\theta)$. To this end it is worth noting that because $\vartheta\neq\theta$ except at the extremes $\vartheta=\theta=0$ and $\vartheta=\theta=\pi/2$, it is impossible to derive reliable values of $\vartheta=\vartheta(r,\theta)$ from Eq.~(\ref{ctau}) if the stellar geometry, i.e. $r=r(\theta)$, is not determined consistently with the rotation law on which depends $\mathscr{T}(\theta)$.\par
   We have calculated  GDF numerically using Eq.~(\ref{f_th}), where $r(\theta)$ is the radius-vector that describes the stellar surface. The function $\vartheta(\theta)$ is determined with Eqs.~(\ref{tau}) and (\ref{ctau}) by interpolation, where $\mathscr{T}(\vartheta)$ at a given $\theta$ is taken as a linear function of $\ln\theta$. For values of $\vartheta$ when $\theta\to\pi/2$ we consider $\mathscr{T}(\vartheta)$ as a linear function of its expression for $\alpha=0$, so that $\vartheta$ is derived by iterating Eq.~(\ref{limtau}).\par
   The behaviour of the function $f(\theta)$ calculated for $\Upsilon=\cos^k\theta$ $\eta=0.2$, 0.8, and for several values of $\alpha$ and $k$ is shown in Fig.~\ref{fig_frt}. In this figure the function $f(\theta)$ calculated with
$\Upsilon=\sin^k\theta$ and $\alpha'=$ $-\alpha/(1+\alpha)$ is also shown. It shows that $f[\cos^k\theta,\alpha]\neq$ $f[\sin^k\theta,\alpha'(\alpha)]$, except for $k=2$ where $f(\theta)$ is the same for both $\Upsilon(\theta)$ functions. The function $f(\theta)$ obtained for $\alpha=0.0$ (green curves in Fig.~\ref{fig_frt}) corresponds to the solution previously obtained by \citet{elr11}.\par

\begin{figure*}[tbp] 
\center\includegraphics[scale=0.9]{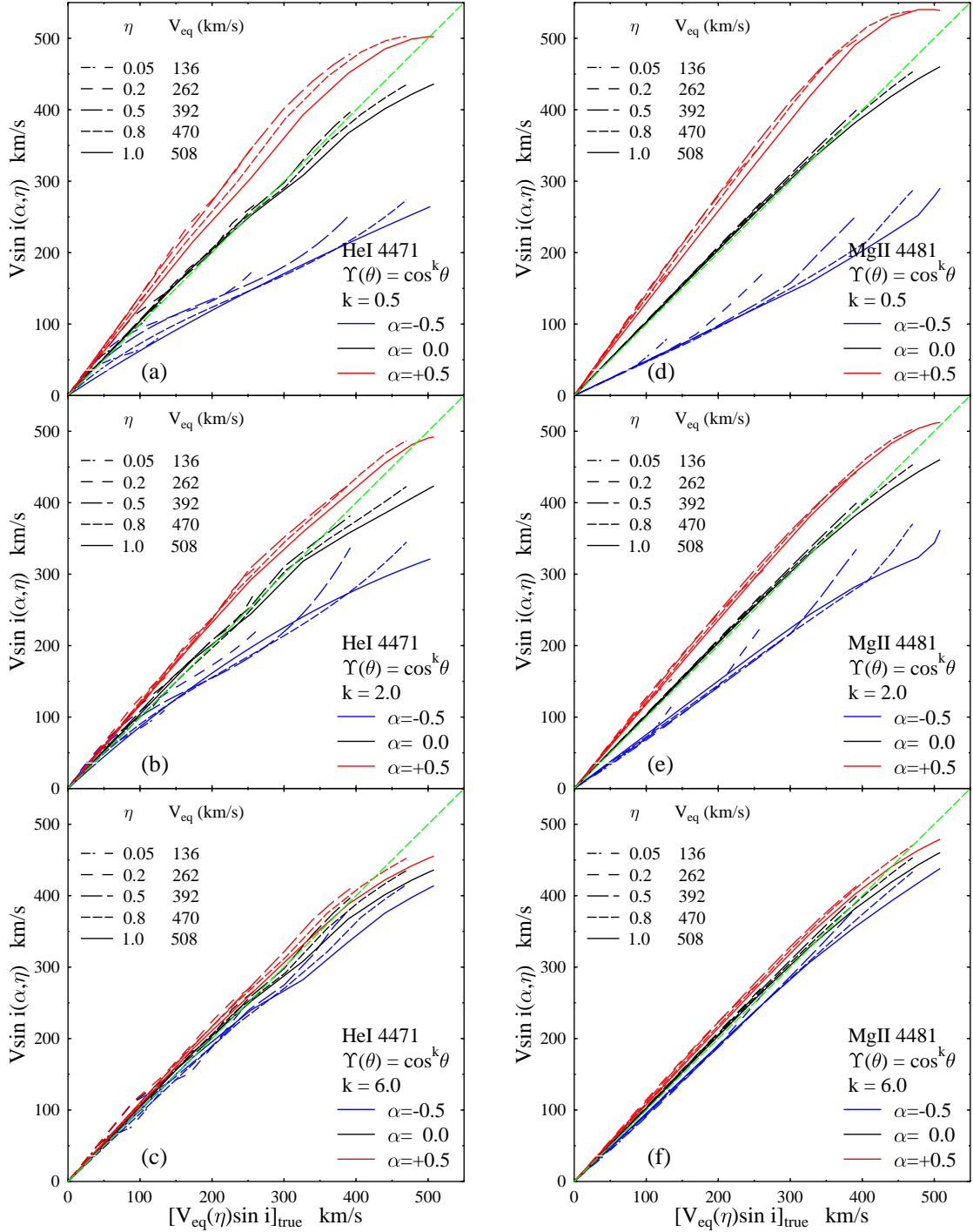} 
\caption{\label{vsini_c} (a), (b) and (c): Parameters $V\!\sin i$ determined with the FT method using the \ion{He}{i} 4471 line produced in atmospheres having a rotation law given by Eq.~\ref{maund} with $\Upsilon=\cos^k\theta$ and characterized by different values of $k$ and $\alpha$, against the actual $V_{\rm eq}\!\sin i$ in stars having $M=9M_{\odot}$ and $t/t_{\rm MS}=0.6$, and rotating with several equatorial ratios $\eta$. (d), (e) and (f): Same for the \ion{Mg}{ii} 4481 line.} 
\end{figure*}  

\begin{figure*}[tbp]
\center\includegraphics[scale=0.9]{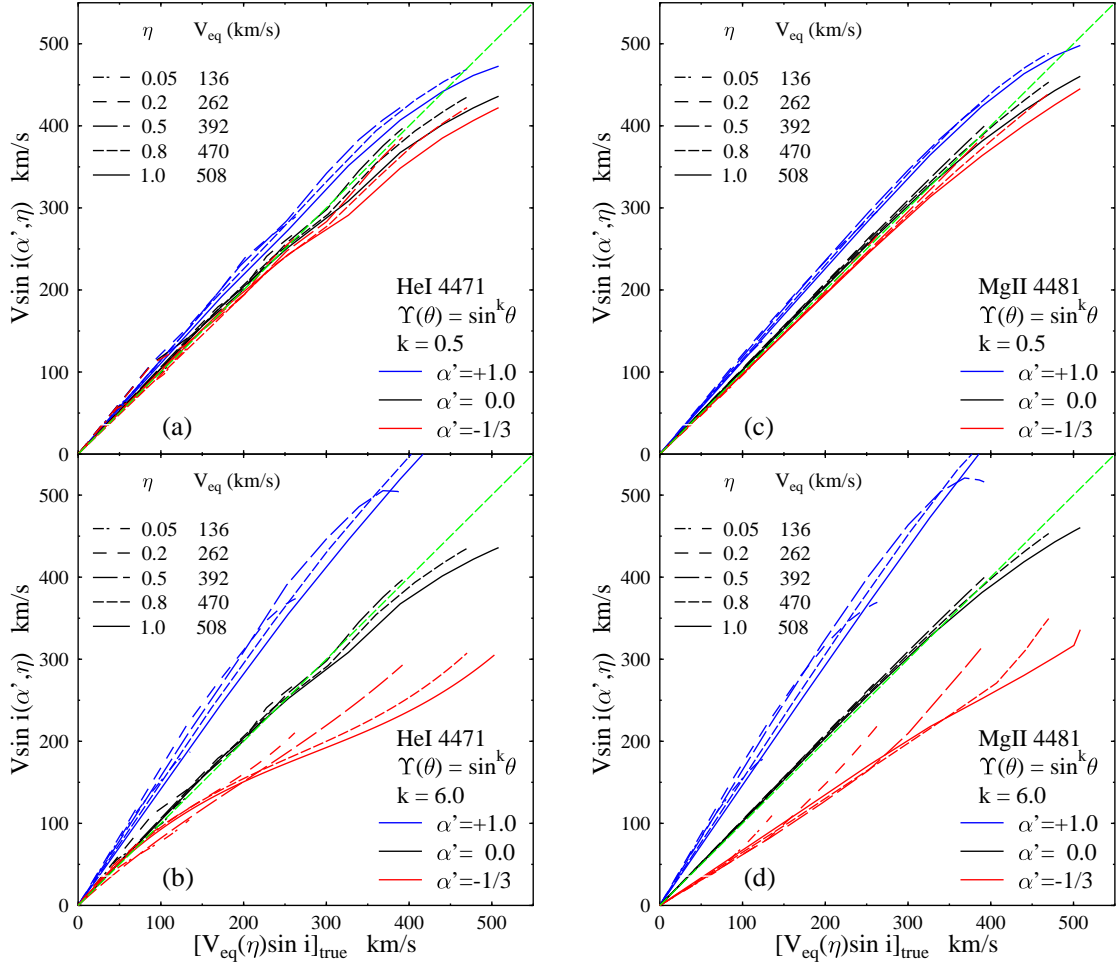}
\caption{\label{vsini_s} Similar as Fig.~\ref{vsini_c} but for $\Upsilon=\sin^k\theta$. The cases with $k=2$ are omitted as they are identical to those in Fig.~\ref{vsini_c}.}
\end{figure*} 

    From Eq.~(\ref{char}) we can derive the limits of $f(r,\theta)$ when $\theta\to0$ and $\theta\to\pi/2$

\begin{eqnarray}
\begin{array}{rcl}
\displaystyle \!\!\!\lim_{\theta\to0}f = f^{\rm pole}\!\!&=& \exp\left[\frac{2}{3}\eta\left(\frac{R_{\rm p}}{R_{\rm e}}\right)^3\left(\frac{\Omega_{\rm p}}{\Omega_{\rm e}}\right)^2\right] \\
\displaystyle \lim_{\theta\to\pi/2}f =\!\!f^{\rm eq} &=& (1-\eta)^{-2/3}, 
\end{array}
\label{limts} 
\end{eqnarray}
 
\noindent where $R_{\rm p}$ and $R_{\rm e}$ are the polar and equatorial radii, respectively. $\Omega_{\rm p}$ is the polar angular velocity, which imply that $\Omega_{\rm p}/\Omega_{\rm e}=1+\alpha$ when $\Upsilon(\theta)=\cos^k\theta$ and $\Omega_{\rm p}/\Omega_{\rm e}=(1+\alpha')^{-1}=(1+\alpha)$ for $\Upsilon(\theta)=\sin^k\theta$. The dependence of $f^{\rm pole}$ with $\alpha$ and $k$ is given through the radii ratio $R_{\rm p}/R_{\rm e}$, which is a function of them both and on the function $\Upsilon(\theta)$ chosen. However, $f^{\rm eq}$ depends only on the ratio $\eta$. It may be worth noting that for $\eta=0$, it is $\vartheta\equiv\theta$ so that in Eq.~(\ref{f_th}) the GDF becomes $f(r,\theta)=1$. This also makes that from Eq.~(\ref{f_th}) we have $f^{\rm pole}=f^{\rm eq}=1$.\par
   For a given ratio $\eta$ we obtain the same value $f^{\rm eq}$ whatever the function $\Upsilon(\theta)$. We may wonder then whether the same value for $f^{\rm pole}$ can be derived whatever the chosen function $\Upsilon(\theta)$ provided we use the appropriate reciprocal gravity-darkening parameter, i.e. $\alpha$ or $\alpha'=-\alpha/(1+\alpha)$, as it is apparent form the expression of $f^{\rm pole}$ given in Eq.~(\ref{limts}). Actually, this reciprocity is
valid only for $k=2$. In the next section we show that for $k\neq2$ this apparent reciprocity does not produce the same ratio $R_{\rm p}/R_{\rm e}$, and the geometry of the stellar surface is not the same either. This is due to the fact that $R_{\rm e}/R_{\rm p}=1+\eta\Gamma$, with $\Gamma=\int_0^{\pi/2}[\Omega(\theta)/\Omega_{\rm e}]^2...d\theta$
[see \citet{zor11}] where, as seen from Fig.~\ref{cosak}, $\Omega(\theta)/\Omega_{\rm e}$ depends on $\Upsilon(\theta)$
for $k\neq2$, even if we use $\alpha$ and its reciprocal $\alpha'$ according to each case. Consequently, the behaviour of the resulting GDF, $f(\theta)$, over the interval $0<\theta<\pi/2$ will depend in a strict sense on the chosen $\Upsilon(\theta)$ function if $k\neq2$.\par

\section{The effective temperature as a function of the colatitude}\label{tac}

    The magnitude of the flux-vector in Eq.~(\ref{eq_vz2}) can be rewritten as

\begin{eqnarray}
\begin{array}{l}
\displaystyle F(\theta) = \sigma_{\rm SB}T^4_{\rm eff}(\theta) = \frac{L}{4\pi GM}f(\theta)g_{\rm eff}(\theta),
\end{array}
\label{defbeta1}
\end{eqnarray}

\noindent where $\sigma_{\rm SB}$ is the \v{S}tefan-Boltzmann constant. Using Eqs.~(\ref{geff_2}) and (\ref{fi}) this relation can be translated into

\begin{eqnarray} 
\begin{array}{l}
\displaystyle T_{\rm eff}(\theta) = \langle T\rangle\left[f(\theta)\gamma(\theta)\right]^{1/4}, 
\end{array}
\label{teff}
\end{eqnarray}

\noindent where the radial variable $r=r(\theta)$ was used as the radius-vector representing the stellar surface, so that $\gamma$ becomes a function of $\theta$ only, apart from its dependence on ($\eta,\alpha$). It is obvious that for $\eta=0$ (no rotation), $f(\theta)=\gamma(\theta)=1$ and $T_{\rm eff}(\theta)=\langle{T}\rangle$. The factor 
$\langle{T}\rangle$ is the uniform effective temperature over a sphere of radius $R_{\rm e}(\eta)$ radiating the luminosity $L$

\begin{eqnarray}
\begin{array}{l}
\displaystyle \langle{T}\rangle^4 = \frac{L}{4\pi R^2_{\rm e}\sigma_{\rm SB}}.
\end{array}
\label{teffm}
\end{eqnarray} 
    
   It must be understood that when interpreting observations the quantity $\langle{T}\rangle$ is a free parameter, because the bolometric luminosity $L$ depends on physical conditions that characterize the core of rotating stars. This luminosity depends, among other factors, on the total angular momentum stored in the star and on its internal distribution, which are unknown. In Figs.~\ref{fig_tgeff}(a) to (c) $T_{\rm eff}(\theta,\eta,\alpha)/\langle T\rangle$ is shown as a function of the co-latitude $\theta$ when $\Upsilon(\theta)=\cos^k\theta$, $\eta=0.8$, $k=0.5$, 2.0 and 6.0, and for several values of $\alpha$ ranging from $-0.5$ to $+0.5$. Figures~\ref{fig_tgeff}(d) to (f) show  $T_{\rm eff}(\theta,\eta,\alpha)/\langle T\rangle$ against $\theta$ for $\Upsilon(\theta)=\sin^k\theta$ and ($k,\alpha'$), where $\alpha'=-\alpha/(1+\alpha)$. The green curves correspond to $T_{\rm eff}(\theta)/\langle{T}\rangle$ for $\alpha=\alpha'=0$, previously obtained for rigid rotators by \citet{elr11}. In Fig.~\ref{fig_tgeffep} we plot the effective temperature ratios $T^{\rm eq}_{\rm eff}/T^{\rm p}_{\rm eff}$ as a function of $\eta$ and different values $k$ and
$\alpha$, when $\Upsilon(\theta)=\cos^k\theta$ or $\Upsilon(\theta)=\sin^k\theta$. 

\begin{figure*}[tbp]
\center\includegraphics[scale=0.9]{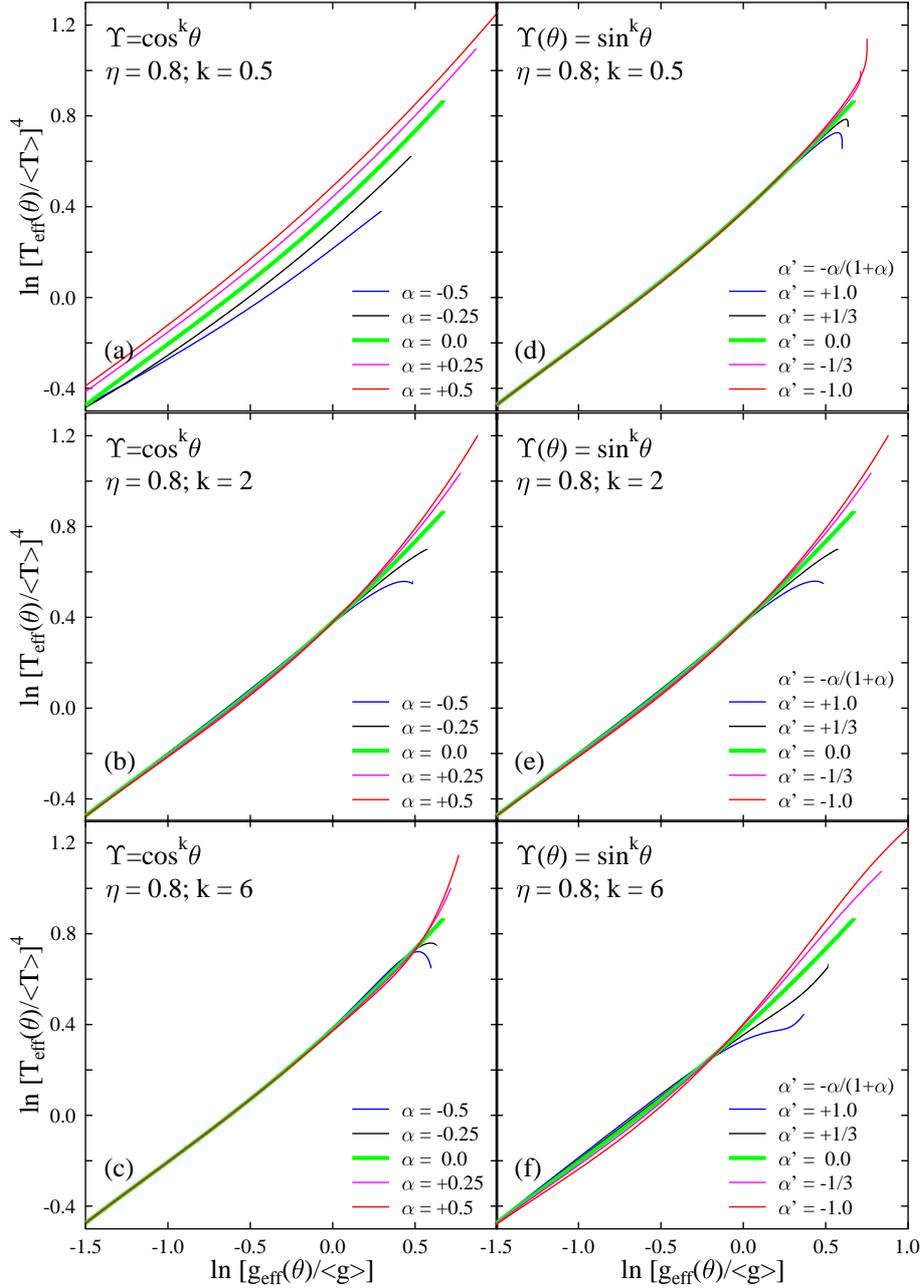} 
\caption{\label{fig_fg} Effective temperature given as $\ln[T_{\rm eff}(\theta)/\langle T\rangle]^4$ against 
$\ln[g_{\rm eff}(\theta)/\langle g\rangle]$ for $\eta=0.8$. (a), (b) and (c) are for $\Upsilon(\theta)=\cos^k\theta$ with $k=0.5$, 2.0 and 6.0, respectively and $\alpha$ ranging from $-0.5$ to $+0.5$. (d), (e) and (f) are for $\Upsilon(\theta)=\sin^k\theta$ with $k=0.5$, 2.0 and 6.0, respectively and $\alpha'(\alpha)$ ranging from $-1.0$ to $+1.0$. In all cases the left-bottom corner corresponds to the equator and the right-upper corner to the pole.}
\end{figure*} 
   
\section{The $V\!\sin i$ parameter}\label{tvsini}  

    $V\!\sin i$ is a parameter that mainly reflects the global broadening of a spectral line. It was shown in \citet{zor17} that this broadening is produced by monochromatic contributions from curved loci of points over the observed stellar hemisphere having the same Doppler displacement. The shape of these curves depend on the differential rotation law and on the inclination angle, while for rigid rotation these curves are straight lines which are independent of the inclination angle. Due to the differentiated sensitivity of spectral lines to the physical formation conditions \citep{thom_83,zor17}, according to the spectral line, the different regions on the observed stellar hemisphere do not contribute with the same efficiency to the rotational broadening because of the non uniform distribution of temperature and gravity. To illustrate this effect, we have calculated \ion{He}{i}\,4471 and \ion{Mg}{ii}\,4481 lines broadened by differential rotators with rotation laws given by Eq.~(\ref{maund}) and both forms of $\Upsilon(\theta)$ given in Eq.~(\ref{maund_l}). For this exercise we used a model star with mass $M=9M_{\odot}$ and fractional age $t/t_{\rm MS}=0.6$ ($t_{\rm MS}$ is the time a rotating star spends in the main sequence evolutionary phase). Rotation laws are characterized by several parameters $k$, $\alpha$, and different equatorial velocities parametrized according to $\eta$. Applying the classic Fourier transform method we determined the $V\!\sin i$ parameters shown in Figs.~\ref{vsini_c} and \ref{vsini_s} where we compare the resulting $V_{\rm eq}\!\sin i$ with the actual input $V_{\rm eq}\!\sin i$ corresponding to the given values of $\eta$ and inclination angles. We note that deviations from the ``identity relation" ($y=x$ line) also exist for rigid rotators. They are due to the gravitational darkening effect, which  is not taken into account in the classic rigid-rotation broadening function $G_{\rm R}(\lambda)$ \citep{gray2}.\par 
    Because it has become of common use today, the classic method based on the FT was employed to determine the $V\!\sin i$. This method assumes, however, that the observed rotationally broadened line profile 
$\mathscr{F}(\lambda)$ can be represented as the convolution with an analytical ``rotation broadening-function" $G_{\rm R}(\lambda)$ of a flux line-profile $F(\lambda)$ emitted by a non-rotating star having uniform effective temperature and gravity. The specific intensity $I(\lambda,\mu)$ contributing to the observed line flux $F(\lambda)$ is thus the same over the entire stellar disc, which obviously does not occur in rotating stars. The derivation of an analytical expression for $G_{\rm R}(\lambda)$ that is independent of the inclination angle requires that the star be spherical and behaves as a rigid rotator \citep{gray1,gray2}, i.e. the monochromatic Doppler displacements are produced over straight strips of constant radial velocities. The $V\!\sin i$ is then obtained by comparing the zeros of the flux line-profile FT $\mathscr{F}(\lambda)$ with the corresponding zeros of the FT of the function 
$G_{\rm R}(\lambda)$. \par
    In \citet{zor17} it has been shown that depending on the spectral line and on the value of $\alpha$, the FT of lines may have unusual shapes, making the zeros difficult to identify and to interpret. This difficultly is a direct consequence of the inconsistency raised by the use an analytic expression $G_{\rm R}(\lambda)$ adapted for uniform
rigid rotators, which does not take into account: a) the actual shape of curves of constant radial velocity contributing to the individual rotational Doppler-displacements, and b) the non uniformity of the effective temperature and surface effective gravity determining the local monochromatic specific intensities $I(\lambda,\mu)$. \par 
    Figures~\ref{vsini_c} and \ref{vsini_s} reveal that the $V\!\sin i$ obtained using the FT-method can be overestimated when $\alpha>0$, and underestimated for $\alpha<0$, as compared to its value for rigid rotators. We notice that for $\Upsilon(\theta)=\cos^k\theta$ and a given $|\alpha|$ the deviations are larger when $\alpha<0$ than
for $\alpha>0$. For $\Upsilon(\theta)=\sin^k\theta$ the deviations reverse according to the $\alpha$-corresponding reciprocal parameter $\alpha'$. Finally, according to the spectral line used, the obtained $V\!\sin i$ is not exactly the same. To apprehend what can be the possible deviations that the current $V\!\sin i$ determinations may incur, let us recall that in the Sun $\alpha\simeq-0.3$ and $\Upsilon(\theta)=\cos^2\theta$, and that predictions made by \cite{rieut13} for rapid rotators of masses $2\lesssim M\lesssim4M_{\odot}$ foresee $\alpha\gtrsim-0.2$. Facing the likely possibility that stars are differential rotators, these results warn us that the current interpretations based on the rigid rotation hypothesis could be misleading. However, detailed modelling of spectral lines can be made today to account for the noted differences in the $V\!\sin i$ obtained from different spectral lines. They could help us to derive indications on the value of the $\alpha$ parameter. \par 
   Needless to say that the classic method of determining $V\!\sin i$ based on the FWHM (full-width at half-maximum) produces the same type of deviations, because the comparisons are made also with the rotational broadening of spectral lines produced by classic spherical rigid rotators.\par  

\begin{figure*}[tbp] 
\center\includegraphics[scale=0.9]{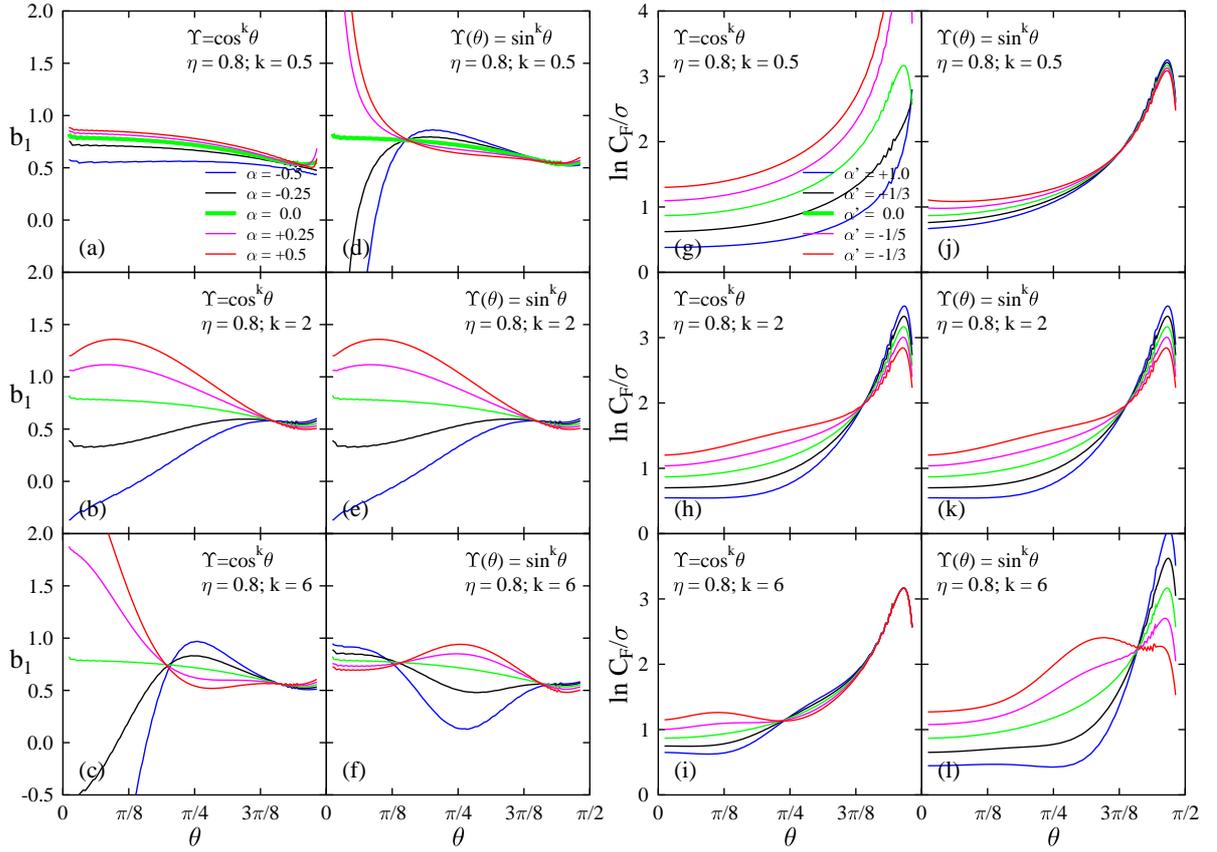} 
\caption{\label{fig_b1} (a), (b) and (c): ``Gravity-darkening slope" $b_1(\overline{\theta})$ against $\theta$ defined in Eq.~(\ref{defpsb2}) for $\Upsilon(\theta)=\cos^k\theta$ with $k=0.5$, 2.0, 6.0, and $\alpha=-0.5$, $-0.25$, 0.0 (green lines), +0.25 and +0.5; (d), (e) and (f): Same, but for $\Upsilon(\theta)=sin^k\theta$ and $\alpha'=+1.0$, 
$+1/3$, 0.0 (green lines), $-1/5$ and $-1/3$; (g), (h) and (i): as from (a) to (c), but for the term $C_{\rm F}(\overline{\theta})$; (j), (k) and (l): as from (d) to (f), but for the term $C_{\rm F}(\overline{\theta})$.}
\end{figure*} 

\section{The gravity-darkening exponent}\label{gde} 

    The ``gravity-darkening exponent" was introduced with the purpose of representing in a simple way the distribution of the local effective temperature $T_{\rm eff}(\theta)$ over a rotating stellar surface. As noted in Sect.\ref{intro}, the known von Zeipel's relation that holds for barotropic objects

\begin{eqnarray}
\begin{array}{l}
\displaystyle \sigma_{\rm SB}T^4_{\rm eff}(\theta) = C_{\rm F}g^{\beta_1}_{\rm eff}(\theta),
\end{array}
\label{defbeta0}
\end{eqnarray}

\noindent is characterized by the constant gravity-darkening exponent $\beta_1$, and the constant $C_{\rm F}$ which depends on the gravity-rotation equipotential of the stellar surface. The notation $\beta_1$ is used here to distinguish it from the exponent $\beta=\beta_1/4$ that is also
frequently employed to characterize the gravity-darkening effect. In the following section, we will comment on some drawbacks of using Eq.~(\ref{defbeta0}).\par 

\begin{figure*}[tbp] \center\includegraphics[scale=1.0]{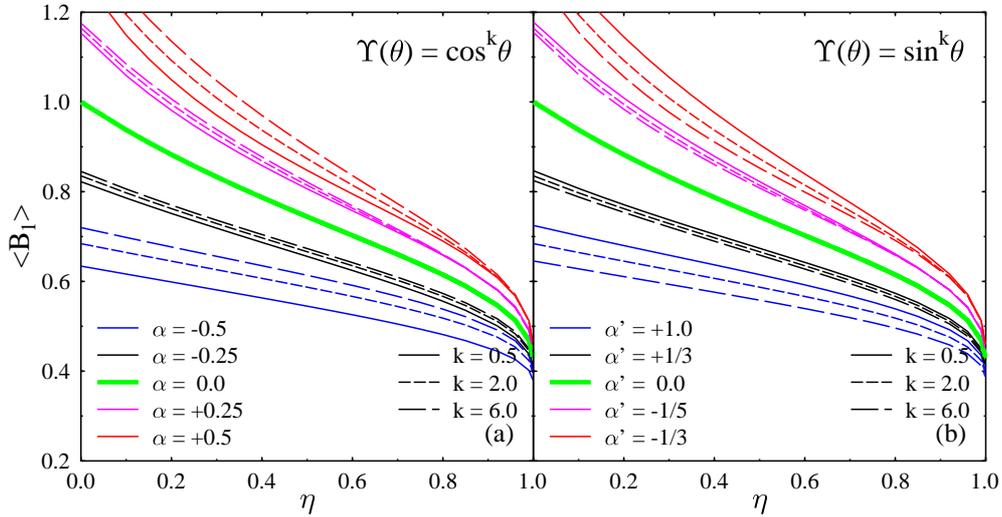}
\caption{\label{fig_b1rieu} Average gravitational-darkening slope $\langle{B_1}\rangle$ given by Eq.~(\ref{gd_co4}) as function of $\eta$, calculated respectively for: (a) $\Upsilon(\theta)=\cos^k\theta$; (b) $\sin^k\theta$. The colours identify the differential rotation parameter $\alpha$, while $k$ is identified using different line-styles.}
\end{figure*}

\subsection{Dependence of the gravity-darkening exponent with colatitude}\label{b1theta}

   Up to now, the relation in Eq.~(\ref{defbeta0}) has been used to interpret the observations of rapidly rotating stars regardless of whether they are barotropic or baroclinic. However, $\beta_1$ and $C_{\rm F}$ cannot be assumed independent of the colatitude-angle $\theta$ even in barotropic model-stars having atmospheres in radiative equilibrium
\citep{elr11,rieut15}.\par 
   To inquire on the effects that a surface differential rotation can induce on $\beta_1$ and $C_{\rm F}$, 
Fig.~\ref{fig_fg} shows the relation between $\ln T_{\rm eff}(\theta)$ calculated with Eq.~(\ref{teff}) against 
$\ln g_{\rm eff}(\theta)$ using both forms of $\Upsilon(\theta)$ in Eq.~(\ref{maund_l}). The model-stars are assumed to rotate with $\eta=0.8$ with several values of $\alpha$ and $k$. Obviously, similar behaviours are noted for other values of $\eta$. \par
   Because the curves in Fig.~\ref{fig_fg} are not straight lines, neither $C_{\rm F}$ nor $\beta_1$ can be considered as genuine constants, even though a quasi-linear relation between $\ln T_{\rm eff}(\theta)$ and $\ln g_{\rm eff}(\theta)$ seems to exist over some intervals of $\theta$. To emphasize more clearly the lack of constancy of $\beta_1$
and $C_{\rm F}$, we define a pseudo-exponent or ``gravity-darkening slope" $b_1(\theta)$ as

\begin{eqnarray} 
\begin{array}{l} 
\displaystyle \ln T^4_{\rm eff}(\theta) = b_1(\theta)\ln g_{\rm eff}(\theta)+\ln[C_{\rm F}(\theta)/\sigma_{\rm SB},] 
\label{defpsb1} 
\end{array} 
\end{eqnarray}

\noindent and consider that between two successive, close discrete points $\theta_i$ and $\theta_{i+1}$, $b_1(\overline{\theta})$ and $C_{\rm F}(\overline{\theta})$ can be assumed constant; i.e.:

\begin{equation}	
\displaystyle b_1(\overline{\theta_i}) = 4\frac{\ln\left[T_{\rm eff}(\theta_{i+1})/T_{\rm eff}(\theta_i)\right]}{\ln \left[g_{\rm eff}(\theta_{i+1})/g_{\rm eff}(\theta_i)\right]}; \ \ \ \overline{\theta_i} = (\theta_i+\theta_{i+1})/2.
\label{defpsb2}
\end{equation}

   Inserting the values $b_1(\overline{\theta_i})$ in Eq.~(\ref{defpsb1}), the corresponding $\ln C_{\rm F}(\overline{\theta_i})$ can be estimated. The inferred quantities $b_1(\overline{\theta_i})$ and $\ln C_{\rm F}(\overline{\theta_i})$ are shown in Fig.~\ref{fig_b1}. They clearly demonstrate that both $b_1$ and $C_{\rm F}$ are dependent on $\theta$ and that consequently they cannot be used as constants in Eq.~(\ref{defbeta0}).

\subsection{The observed gravity-darkening slope}\label{db1i}

\subsubsection{Neglecting the viewing angle}\label{db1i_noi}

   Replacing in Eq.~(\ref{defpsb1}) the expressions for $g_{\rm eff}$ and $T_{\rm eff}$ given respectively by Eqs.~(\ref{geff_2}) and (\ref{teff}), and imposing that $b_1$ and $C_{\rm F}$ do not depend on $\theta$, it follows that

\begin{eqnarray} 
\begin{array}{l}
\displaystyle B_1(\theta) = 1+ \frac{\ln[f(\theta)/f^{\rm pole}]}{\ln[\gamma(\theta)/\gamma^{\rm pole}]},
\end{array}
\label{defb1}
\end{eqnarray}

\noindent where $f^{\rm pole}$ and $\gamma^{\rm pole}$ are the values of the functions $f(\theta)$ and $\gamma(\theta)$ in $\theta=0$. The function $B_1(\theta)$ behaves in a similar way as $b_1(\theta)$, so that it is needless to reproduce it graphically. Instead, we can replace $f(\theta)$ and $\gamma(\theta)$ in Eq.~(\ref{defb1}) by their values
at $\theta=\pi/2$ [see Eq.~(\ref{limts})] to obtain a kind of average $\langle B_1(\theta)\rangle$ that takes the form

\begin{eqnarray} 
\begin{array}{l} 
\displaystyle \langle{B_1}\rangle = 1-\frac{2}{3}\left[\frac{\ln(1-\eta)+\eta(R_{\rm p}/R_{\rm
e})^3(\Omega_{\rm p}/\Omega_{\rm e})^2}{\ln[(1-\eta)+2\ln(R_{\rm p}/R_{\rm
e})}\right], 
\end{array} 
\label{gd_co4} 
\end{eqnarray}

\noindent which is a generalized version of $\langle B_1\rangle$ obtained by \citet{elr11} for rigid rotators. In Fig.~\ref{fig_b1rieu}, $\langle B_1\rangle$ is shown as a function of $\eta$ for $-0.5\leq\alpha\leq0.5$, $k=0.5$, 2, 6, and for both functions $\Upsilon(\theta)$. The expression of $\langle B_1\rangle$ for $\alpha=0$ was sometimes used in the literature to compare the exponents $b_1$ derived from interferometric observations. \par

\begin{figure}[tbp] 
\center\includegraphics[scale=0.9]{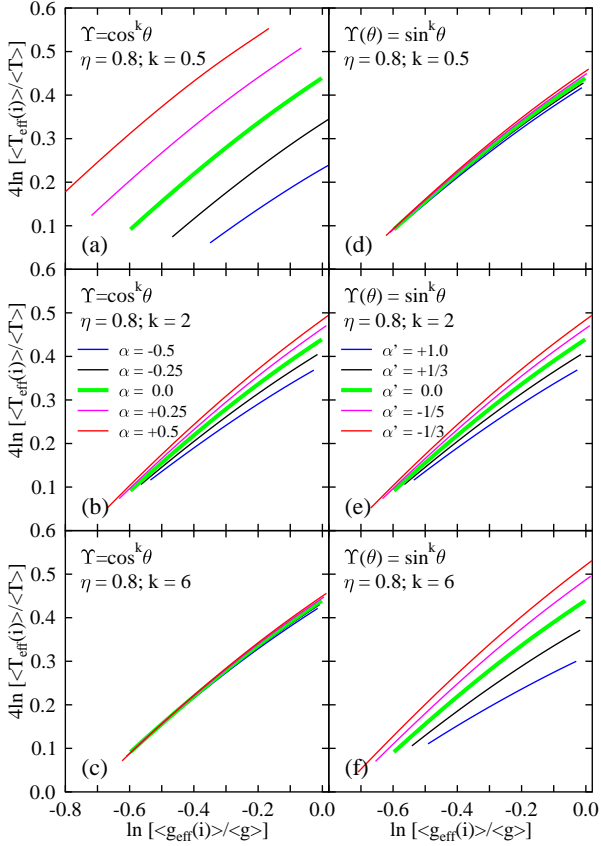}
\caption{\label{fig_fgi} Relation between the apparent $4\ln[T_{\rm eff}(i)/\langle T\rangle]$ and $\ln[g_{\rm eff}(i)/\rangle\langle g\rangle]$; (a), (b) and (c) for $\Upsilon(\theta)=\cos^k\theta$ with $\alpha$ from $-0.5$ to +0.5; (d), (e) and (f) for $\sin^k\theta$ and $\alpha'=$ $-\alpha/(1=\alpha)$. All curves are for $\eta=0.8$, with $k=0.5$, 2.0, 6.0. The colour-codes according to $\alpha$ and $\alpha'$ indicated in the respective blocks for $k=2$ hold for all curves having the same $\Upsilon(\theta)$. The equator-on seeing directions ($i\to\pi/2$) are in the lower left corner of each block, while the polar-on directions ($i\to0$) are in the upper right corner.}
\end{figure} 

\subsubsection{Considering the viewing angle}\label{db1i_withi}

    Any parameter describing a physical property of a rotating stellar atmosphere that depends on the co-latitude 
$\theta$, automatically becomes a function of the viewing angle $i$ when it is determined from observational data of angularly unresolved stars. This must also happen to the gravity-darkening slope given by Eq.~(\ref{defb1}). Such a
quantity should thus depend on the apparent average effective temperature and gravity that characterize the observed stellar hemisphere, which are named hereafter as $\langle{T}_{\rm eff}(i)\rangle$ and $\langle{g}_{\rm eff}(i)\rangle$, respectively. \par
   Let us define $\langle{T}_{\rm eff}(i)\rangle$ and $\langle{g}_{\rm eff}(i)\rangle$ so as to obtain a slope that might be assimilated with an empirically derived gravity-darkening slope. The apparent effective temperature $\langle{T}_{\rm eff}(i)\rangle$ can be written as

\begin{eqnarray}
\begin{array}{l}
\displaystyle \langle T_{\rm eff}^4(i)\rangle = \frac{L(i)}{\sigma_{\rm SB}\mathcal{S}(i)},
\end{array}
\label{teff_i}
\end{eqnarray}  

\begin{figure*}[tbp] 
\center\includegraphics[scale=0.9]{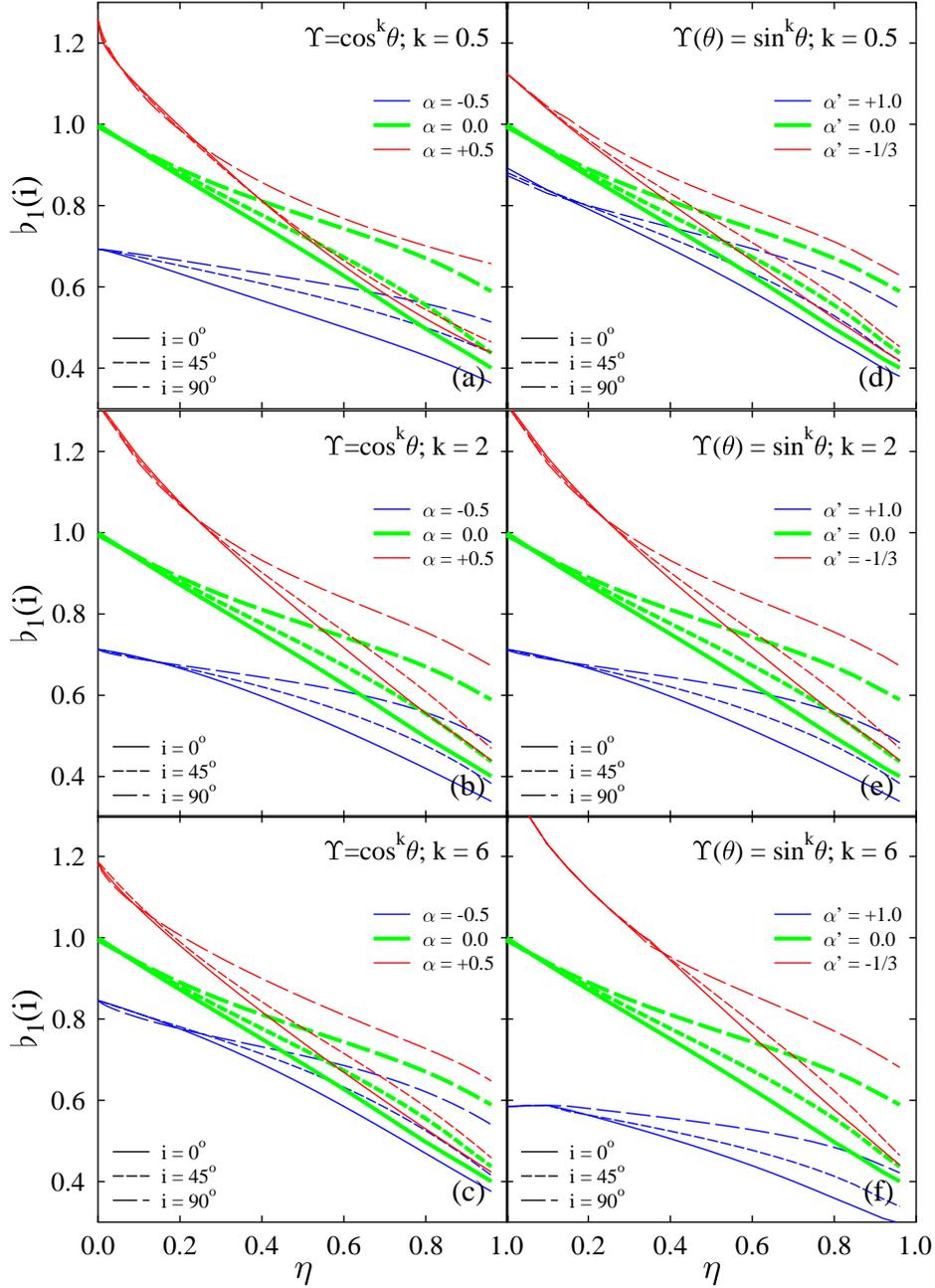}
\caption{\label{fig_b1ie} Gravitational-darkening slope $\flat_1(i)$ against the rotational ratio $\eta$. The colours identify the values of the differential parameter $\alpha$, and the line-styles are for the inclination angles. In (a) to (c) the slope $\flat_1(i)$ was calculated with $\Upsilon(\theta)=\cos^k\theta$ and from (d) to (f) $b_1(i)$ was obtained with $\Upsilon(\theta)=\sin^k\theta$. As throughout in this work, rigid rotation ($\alpha=0$) is identified with green lines.}
\end{figure*}  

\noindent where $\mathcal{S}(i)$ is the effective emitting area of the $i-$dependent rotationally deformed stellar hemisphere. In this expression, the apparent bolometric luminosity $L(i)$ emitted by the stellar disc projected towards the observer is given by \citep{coll65,maedp70,coll73}

\begin{eqnarray}
\begin{array}{l}
\displaystyle L(i) = 4\pi R^2_{\rm
e}\oint_{S(i)}I(\mu)\left(\frac{\hat{\rm\bf
n}\cdot\hat{\boldsymbol\iota}}{\hat{\rm\bf n}\cdot\hat{\vec{e}}_r}\right)r^2(\theta)\sin\theta\,d\theta d\phi,
\end{array}
\label{lum_i}
\end{eqnarray}
 
\noindent where $\hat{\rm\bf n}=$ $-\vec{g}_{\rm eff}/g_{\rm eff}$ is the unit vector normal to the stellar surface; 
$\hat{\boldsymbol\iota}$ is the unit vector directed towards the observer; $\hat{\vec{e}}_r$ is the unit vector associated with the $r-$spherical coordinate; $S(i)$ indicates the observed region of the stellar hemisphere over which
the integration is carried out. $\mu$ is the directional cosine $\cos(\hat{\rm\bf n},\hat{\boldsymbol\iota})=$ 
$\hat{\rm\bf n}\cdot\hat{\boldsymbol\iota}$; $I(\mu)$ is the bolometric specific intensity of the radiation for which we use a quadratic form

\begin{eqnarray}
\begin{array}{l}
\displaystyle I(\mu) = I_1\left[1-\epsilon_1(1-\mu)-\epsilon_2(1-\mu^2)\right],
\end{array}
\label{i_bol1}
\end{eqnarray}  

\noindent whose limb-darkening coefficients $\epsilon_{1,2}$ have been interpolated in the tables given by \citet{clar00} for each pair of local fundamental parameters [$T_{\rm eff}(\theta),g_{\rm eff}(\theta)$]. Knowing that for each point over the stellar surface we have $|\vec{F}|=F(\theta)$ $=2\pi\int_0^1I(\mu)\mu\,d\mu$, the intensity $I_1=I(\mu=1)$ can be written as

\begin{eqnarray}
\begin{array}{l}
\displaystyle I_1(\theta)=\frac{6}{\pi}\left[\frac{F(\theta)}{6-2\epsilon_1-3\epsilon_2}\right],
\end{array}
\label{i_bol2}
\end{eqnarray} 

\noindent which depends on $f(\theta)$ and $\gamma(\theta)$ through $F(\theta)$ given in Eq.~(\ref{defbeta1}).\par 
    Consistently with the definition of $\langle{T}_{\rm eff}(i)\rangle$, we can use Eq.~(\ref{defbeta1}) to estimate 
$\langle{g}_{\rm eff}(i)\rangle$ using the expression

\begin{eqnarray} 
\begin{array}{l} 
\displaystyle \langle T^4_{\rm eff}(i)\rangle = \frac{L}{4\pi GM\sigma_{\rm SB}}\langle f g_{\rm eff}\rangle_i, 
\end{array} 
\label{geff_i} 
\end{eqnarray}

\noindent that with Eqs.~(\ref{teff_i}) and (\ref{lum_i}) leads to

\begin{eqnarray}
\begin{array}{l}
\displaystyle \langle g_{\rm eff}(i)\rangle = \frac{\langle f\,g_{\rm eff}\rangle_i}{\langle f\rangle_i},
\end{array}
\label{geff_ii}
\end{eqnarray}
 
\noindent where $\langle{f}\rangle_i$ is given by

\begin{eqnarray}
\begin{array}{rcl}
\displaystyle \langle f\rangle_i & = & \displaystyle
\left.\oint_{S(i)}f(\theta)\phi(\mu)\left(\frac{\hat{\rm\bf
n}\cdot\hat{\boldsymbol\iota}}{\hat{\rm\bf n}\cdot\hat{\vec{e}}_r}\right)r^2(\theta)\sin\theta\,d\theta d\phi\middle/\mathcal{S}(i)\right. \\
\displaystyle \phi(\mu) & = & \displaystyle \pi^{-1}[1\!-\!\epsilon_1(1\!-\!\mu)\!-\!\epsilon_2(1\!-\!\mu^2)]/(1\!-\!\epsilon_1/3\!-\!\epsilon_2/2).
\end{array}
\label{geff_iii}
\end{eqnarray}

   For the numerical estimates of the apparent $\langle{T}_{\rm eff}(i)\rangle$ and $\langle{g}_{\rm eff}(i)\rangle$ the knowledge of absolute values of $T_{\rm eff}(\theta)$ and $g_{\rm eff}(\theta)$ are required. To get a rough insight on their amplitude variation, we have assumed model-stars with mass $M=9M_{\odot}$ and fractional age
$t/t_{\rm MS}=0.6$ ($t_{\rm MS}$ is the time spent by a rotating star in the main sequence phase). The bolometric luminosity $L$ was calculated with the rotationally induced mass-compensation effect parametrized as in \citet{frem05}, where it is assumed that the stellar core rotates rigidly. The equatorial radii ratio $R_{\rm e}/R_{\rm o}$ 
($R_{\rm o}$ is the equatorial radius of a parent non-rotating object having the same mass) is given by the models of \citet{zor11}.\par
   The relations between $\ln[\langle{T}_{\rm eff}(i)\rangle/\langle{T}\rangle]$ and $\ln[\langle{g}_{\rm
eff}(i)\rangle/\langle{g}\rangle]$ shown in Fig.~\ref{fig_fgi} are calculated using $\Upsilon(\theta)=\cos^k\theta$ and $\sin^k\theta$, for $\eta=0.8$, $k=0.5$, 2.0, 6.0, and $-0.5\leq\alpha\leq-0.5$. While Fig.~\ref{fig_fg} represents the variation of $T_{\rm eff}$ against $g_{\rm eff}$ over the stellar surface, Fig.~\ref{fig_fgi} represents these parameters integrated over a stellar surface seen according to different viewing angles. The upper right corner corresponds to pole-on seeing directions and the lower left corners to the equator-on directions. For a given geometrical deformation of stars parameterized here with the ratio $\eta$, it turns out that the resolution among these curves according to the differential parameter $\alpha$ is a strong function of the power $k$ used in 
$\Upsilon(\theta)$.\par
   From Fig.~\ref{fig_fgi} it seems legitimate to assume relations of the type

\begin{eqnarray} 
\begin{array}{l}
\displaystyle \ln\langle{T}^4_{\rm eff}(i)\rangle = \flat(i)\ln\langle g_{\rm eff}(i)\rangle + \langle A(i)\rangle
\end{array}
\label{defb1i}
\end{eqnarray}  

\noindent for each $[\Upsilon,k,\alpha]$ so as to represent the dependence of the slope $\flat(i)$ with the stellar inclination, as we did for $b_1(\theta)$ with $\theta$, i.e. by taking here $\flat(i)$ and $\langle A(i)\rangle$ as constants between two successive inclinations $i_{j}$ and $i_{j+1}$ in a given discretization of this variable in $0\degr\leq i\leq90\degr$.  The values of $\flat(i)$ obtained for three inclination angles: $i=0^\circ$, $45^\circ$ and $90^\circ$ against the ratio $\eta$ are shown in Fig.~\ref{fig_b1ie}. Figures~\ref{fig_b1ie}a,b,c are for $\Upsilon=\cos^k\theta$ with $k=0.5$, 2 and 6, while Fig~\ref{fig_b1ie}d,e,f are for $\Upsilon=\sin^k\theta$ and the same
exponents $k$. In these figures we note that the $\flat(i)$ curves have a global behaviour that is similar to 
$\langle{B_1}\rangle$ in Fig.~\ref{fig_b1rieu}, except for their spread as a function of $i$. \par
\begin{figure}[tbp] \center\includegraphics[scale=1.0]{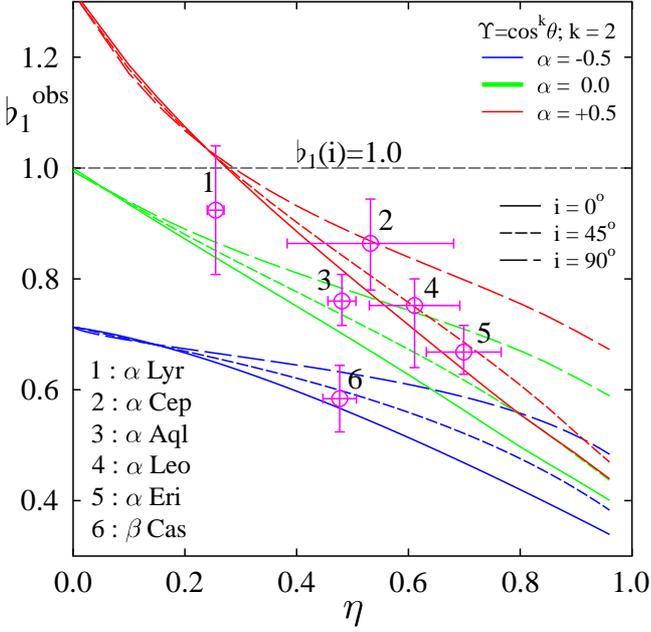}
\caption{\label{fig_b1obs} Comparison of the empirical gravitational-darkening slope $\flat_1^{\rm obs}$ \citep[see
ref. in][]{souz14} with the curves representing the $\flat_1(i)$ of rotators with $k=2$, $\Upsilon(\theta)=\cos\theta$, $\alpha=-0.5$, 0.0 (green curves) and $\alpha=+0.5$, and inclination angles $i=0\degr$, $45\degr$ and $90\degr$ (dentified with different line styles). The dashed black dashed line is for $\flat_1(i)=1$.} 
\end{figure}

   In Fig.~\ref{fig_b1obs} we show the values of $b_1^{\rm obs}$ determined with interferometric data \citep[see][]{souz14}, where the interpretation of observations was done using Eq.~(\ref{defbeta0}). These points have been originally given as a function of the stellar flattening $\varepsilon=1-R_{\rm p}/R_{\rm e}$. We have converted 
$\varepsilon$ into $\eta$ considering, as the authors themselves did, that stars are rigid rotators, i.e. 
$R_{\rm e}/R_{\rm p}=1+\eta/2$, which imply that $\eta=2\varepsilon/(1-\varepsilon)$. In this diagram the curves
$\flat_1(i)$ are reproduced for $\Upsilon(\theta)=\cos^k\theta$ with $k=2$ and different values of $\alpha$, in particular those suited for rigid rotators, i.e.  $\alpha=0$ (green curves). In Sect.~\ref{disc_concl} we discuss the reason of using $k=2$. Taking into account the observational uncertainties, we can conclude that half of the observed points are located outside the zone limited by $\flat_1(i=0)$ and $\flat_1(i=\pi/2)$ of rigid rotators. Two of them are likely situated in the region of $\alpha>0$, while there is one in the zone of $\alpha<0$, which indicates that other solutions than rigid rotation might also suit these objects.\par
   However, according to the diagrams in Fig.~\ref{fig_b1ie} it clearly appears that there is not a unique way to characterize the studied stars with $\beta_1$ as rotators, since many solutions can be envisioned in terms of 
$[\Upsilon,k,\alpha]$. On the other hand, there is a strong concern with the interpretation of the observed values 
$\flat_1^{\rm obs}$, because $\beta_1$ and $C_{\rm F}$ were considered as constants in the reduction of data. Due to these ambiguities, it would perhaps be wiser to abandon the parameter $\beta_1$ to model fluxes and rather concentrate on  determining them using the differential rotation parameter $\alpha$ and the inclination angle $i$ \citep{domic_2004}.\par

\section{Discussion and conclusion}\label{disc_concl}  

   Continuing the work initiated by \citet{elr11} to represent the distribution of the effective temperature in rotating stars, where the only requirement is that the outermost layers of the envelope be in radiative equilibrium and the radiation flux vector be dominantly anti-parallel to the effective surface gravity, we have discussed the case where these layers are in differential rotation. We have studied the particular case where the angular velocity is given by Eqs.~(\ref{maund}) and (\ref{maund_l}), and obtained relationships that generalize those previously derived by \citet{elr11} for rigid rotators. As a further step, we have studied the effects of the surface differential rotation on the determination of the $V\!\sin i$ parameter and concluded that strong deviations can be expected. They seem to be large enough to be exploited empirically. We also find that a  constant gravity-darkening exponent is inappropriate for describing stars as rotators. In fact, as this exponent is a function of the stellar co-latitude \cite[][]{rieut15}, it becomes a function of the viewing angle that induces ambiguities in the interpretation of observations. The modelling of observed parameters of rotating stars must then be done directly through the gravity-darkening-function (GDF), $f(\theta)$, which describes the distribution of the effective temperature as indicated by Eqs.~(\ref{teff}), (\ref{f_th}), (\ref{ctau}) and (\ref{intau}), for which the stellar geometry must be calculated consistently. According to discussions carried out in this work, the interpretation of observations would then require one to determine the set of quantities $(\eta,\alpha,\Upsilon)$ instead of the simple pair $(\eta,\beta_1)$.\par    
   In the approach where differential rotation is considered, a significant uncertainty remains in the choice of the function $\Upsilon(\theta)$. Indeed, up to now, and except for the Sun, no empirical indication exists on the actual nature of $\Upsilon(\theta)$ in stars. The quantity that needs to be inferred first is a parameter related to the average surface gradient $\langle{d\,\Omega/d\,\theta}\rangle$. To this end, the simplest expression for 
$\Upsilon(\theta)$ could be $2\theta/\pi$. This makes $\alpha$ the unique free parameter characterizing the surface differential rotation, but in this case the calculation of the integral $\mathscr{T}(\theta)$ [Eq.~(\ref{intau})]
has some annoying numerical subtleties. However, we have shown in Sect.~\ref{tsdr} that for the same ratio 
$\Omega_{\rm p}/\Omega_{\rm e}$, no significant differences in the prediction or analysis of observed
parameters may arise if we use only $\Upsilon(\theta)=\cos^k\theta$ or $\sin^{k'}\theta$, because reliable correspondences can be established between $(\alpha,k)$ and $(\alpha',k')$. According to this finding, we can adopt 
$\Upsilon(\theta)=\cos^k\theta$ and show easily that the weighted average gradient over the stellar surface

\begin{eqnarray}   
\begin{array}{l}
\displaystyle \langle d(\Omega/\Omega_o)/d\theta\rangle = \int_0^{\pi/2}[d(\Omega/\Omega_o)/d\theta]\sin\theta 
d\,\theta
\end{array}
\label{weig_om}  
\end{eqnarray}

\noindent is equal to $\langle d(\Omega/\Omega_o)/d\theta\rangle=$ $2\alpha/\pi$ if $k=2.55$. Although nothing prevents us from making numerical models with $k=2.55$, the results for $T_{\rm eff}(\theta)$, $V\!\sin i$ and other parameters, show that no significant differences occur on the predicted or analyzed parameters if we simply adopt $k=2$. Moreover, the representation $\Upsilon(\theta)=\cos^2\theta$ of \cite{maunder_maunder1905} accounts for the solar surface differential rotation. Two first order indications could then be reliably drawn from observations: a) whether the surface differential rotation in stars is significant or not; b) what is its character: accelerated towards the pole or accelerated towards the equator. We note that for some forms of $\Upsilon(\theta,k)$, in particular for $k=2$, analytical expressions of $\mathscr{T}(\theta)$ exist and can help to proceed in a similar way as for rigid rotators with the formulas developed by \citet{elr11}. \par

\begin{acknowledgements} 
We are thankful to the referee for his/her attentive reading of the manuscript and for his/her suggestions to improve the presentation of our results.\par 
\end{acknowledgements}

\bibliographystyle{aa} 
\bibliography{30818}

\begin{appendix}

\onecolumn

\section{The functions $\mathscr{T}(\theta)$} 
\label{tft} 

  The analytical expressions obtained of the function $\mathscr{T}(\theta)$ with $\Upsilon(\theta)= \cos^k(\theta)$ by integrating Eq.~(\ref{dtautt}) for the natural numbers $k=1$, 2, 3 and 4 are:\par
\medskip
\noindent $k=1$\par

\begin{eqnarray}
\begin{array}{rcl} 
\displaystyle \mathscr{T}(\theta) & = & \displaystyle\frac{1}{(1-\alpha^2)^2}
\left\{(1+\alpha^2)\ln\tan\frac{\theta}{2}+2\alpha\ln\left|\frac{1+\alpha\cos\theta}{\sin\theta}\right|+(1-\alpha^2)\left(\frac{\cos\theta}{1+\alpha\cos\theta}\right) \right\} \ \ \ \ \forall \alpha
\end{array}
\label{tau_k1}
\end{eqnarray}

\noindent $k=2$\par

\begin{eqnarray}
\begin{array}{rcl} 
\displaystyle \mathscr{T}(\theta) & = &
\left\{
\begin{array}{ll}
\displaystyle\frac{1}{(1+\alpha)^2}\left[\ln\tan\frac{\theta}{2}+(1-\alpha)\frac{\arctan(\sqrt{\alpha}\cos\theta)}{2\sqrt{\alpha}} \right. + 
\left.\frac{(1+\alpha)}{2}\left(\frac{\cos\theta}{1+\alpha\cos^2\theta}\right)\right] & {\rm for}\ \alpha \geq 0 \\
       \\
\displaystyle \frac{1}{(1-|\alpha|)^2}\left[\ln\tan\frac{\theta}{2}+
\left(\frac{1+|\alpha|}{4\sqrt{|\alpha|}}\right)\ln\left(\frac{1+\sqrt{|\alpha|}\cos\theta}{1-\sqrt{|\alpha|}\cos\theta}\right)\right. + \left.\left(\frac{1-|\alpha|}{2}\right)\left(\frac{\cos\theta}{1-|\alpha|\cos^2\theta} \right)\right] & {\rm for}\ \alpha<0
\end{array}
\right.
\end{array}
\label{tau_k2}
\end{eqnarray}

\noindent $k=3$\par

\begin{eqnarray}
\begin{array}{rcl} 
\displaystyle \mathscr{T}(\theta) & = &
\left\{
\begin{array}{ll}
\displaystyle \frac{1}{(1-\alpha^2)^2}\left\{(1+\alpha^2)\ln\tan\frac{\theta}{2}-2\alpha\ln\sin\theta+\frac{2}{3}\alpha\ln(1\pm u^3)+ \frac{2}{3}[1+2\alpha^2+|\alpha|^{2/3}(2+\alpha^2)]\left(\frac{A}{|\alpha|^{1/3}}\right)\right. + & \\
\displaystyle \left.\frac{2}{3}[1+2\alpha^2-|\alpha|^{2/3}(2+\alpha^2)]\left(\frac{B}{|\alpha|^{1/3}}\right)+\frac{1}{3}\frac{\cos\theta}{(1\pm u^3)}[1-\alpha\cos\theta(1-\alpha\cos\theta)]\right\} & \\
\displaystyle u = |\alpha|^{1/3}\cos\theta; \ \ \ A = \frac{1}{6}\ln\left[\frac{(1\pm u)^2}{1-(\pm)u+u^2}\right]; \ \ \ B = \frac{1}{\sqrt{3}}\left[\arctan\left(\frac{2u-(\pm)1}{\sqrt{3}}\right)\pm\frac{\pi}{6}\right]; & \\
``+"\ \  {\rm for}\ \alpha \geq 0 \ \ \ {\rm and} \ \ \ ``-" \ \ {\rm for}\ \alpha < 0 \\
\end{array}
\right.
\end{array}
\label{tau_k3}
\end{eqnarray}

\noindent $k=4$\par

\begin{eqnarray}
\begin{array}{rcl} 
\displaystyle \mathscr{T}(\theta) & = &
\left\{
\begin{array}{ll}
\displaystyle\frac{1}{(1+\alpha)^2}\left\{\ln\tan\frac{\theta}{2}+\frac{\left[3-\alpha+\sqrt{\alpha}(5+\alpha)\right]}{16\sqrt{2}\alpha^{1/4}}\ln\left[\frac{1+\alpha^{1/4}\cos\theta(\sqrt{2}+\alpha^{1/4}\cos\theta)}{1-\alpha^{1/4}\cos\theta(\sqrt{2}-\alpha^{1/4}\cos\theta)}\right] \right.+ & \\
\displaystyle \left.\frac{\left[3-\alpha-\sqrt{\alpha}(5+\alpha)\right]}{8\sqrt{2}\alpha^{1/4}}\arctan\left(\frac{\sqrt{2}\alpha^{1/4}\cos\theta}{1-\alpha^{1/2}\cos^2\theta}\right)+\frac{(1+\alpha)}{4}\cos\theta\left(\frac{1-\alpha\cos^3\theta}{1+\alpha\cos^4\theta}\right)\right\} & {\rm for}\ \alpha \geq 0 \\
       \\
\displaystyle\frac{1}{(1-|\alpha|)^2}\left\{\ln\tan\frac{\theta}{2}+\frac{\left[3+|\alpha|+\sqrt{|\alpha|}(5-|\alpha|)\right]}{16|\alpha|^{1/4}}\ln\left|\frac{1+|\alpha|^{1/4}\cos\theta}{1-|\alpha|^{1/4}\cos\theta}\right| \right.+ & \\
\displaystyle \left.\frac{\left[3+|\alpha|-\sqrt{|\alpha|}(5-|\alpha|)\right]}{8|\alpha|^{1/4}}\arctan\left(|\alpha|^{1/4}\cos\theta\right)+\frac{(1-|\alpha|)}{4}\cos\theta\left(\frac{1+|\alpha|\cos^3\theta}{1-|\alpha|\cos^4\theta}\right) \right\} & {\rm for}\ \alpha < 0
\end{array}
\right.
\end{array}
\label{tau_k4}
\end{eqnarray}

  The corresponding analytical expressions of $\mathscr{T}(\theta)$ with $\Upsilon(\theta)= \sin^k(\theta)$ by integrating Eq.~(\ref{dtautt}) for the natural numbers $k=1$ and 2, are:\par
\medskip

\noindent $k=1$\par

\begin{eqnarray}
\begin{array}{rcl} 
\displaystyle \mathscr{T}(\theta) & = & \displaystyle \ln\tan\frac{\theta}{2}+ \frac{\cos\theta}{1+\alpha\sin\theta}-\frac{2\alpha}{(1-\alpha^2)^{1/2}}\left\{\arctan\left[\frac{\tan\frac{\theta}{2}+\alpha}{1-\alpha^2)^{1/2}}\right]-\arctan\left(\frac{1+\alpha}{1-\alpha}\right)^{1/2}\right\}; \ \ \ {\rm for} \ \ \alpha^2<1
\end{array}
\label{tau_ks1}
\end{eqnarray}

\noindent $k=2$\par

\begin{eqnarray}
\begin{array}{rcl} 
\displaystyle \mathscr{T}(\theta) & = & \displaystyle \ln\tan\frac{\theta}{2} + \left\{
\begin{array}{ll}
\displaystyle \frac{1}{2}\left(\frac{\cos\theta}{1+\alpha\sin^2\theta}\right)-\frac{1}{4}\frac{(1+2\alpha)}{[\alpha(1+\alpha)]^{1/2}}\ln\left(\frac{1-u}{1+u}\right);  \ \ \ u = \left(\frac{\alpha}{1+\alpha}\right)^{1/2}\!\!\!\!\cos\theta; \ \ \ {\rm for} \ \alpha \geq 0 & \\
\displaystyle \frac{1}{2}\left(\frac{\cos\theta}{1-|\alpha|\sin^2\theta}\right)+\frac{1}{2}\frac{(1+2|\alpha|)}{[|\alpha|(1-|\alpha|)]^{1/2}}\arctan\left[\left(\frac{|\alpha|)}{1-|\alpha|}\right)^{1/2}\!\!\!\!\cos\theta\right] \ \ \ {\rm for} \ \ \alpha < 0 
\end{array}
\right.
\end{array}
\label{tau_ks2}
\end{eqnarray}

    In spite of the formal difference between Eq.~(\ref{tau_k2}) and Eq.~(\ref{tau_ks2}), they are equivalent.
They can be converted to each other knowing that $\sin^2\theta+\cos^k\theta=1$, making the transformation
$\alpha$ into $-\alpha/(1+\alpha)$ and considering the sign of $\alpha$.\par

\section{Limit forms of $f(r,\theta)$}
\label{lef} 

 Integrating Eq.~(\ref{intau}) by parts, we obtain the following expression equivalent to 
Eq.~(\ref{tau})

\begin{eqnarray}
\begin{array}{rcl}
\displaystyle \ln\left[\frac{\tan(\vartheta/2)}{\tan(\theta/2}\right]+\cos\vartheta-\cos\vartheta\left[\frac{1+\alpha\Upsilon(\vartheta)}{1+\alpha\Upsilon(\theta)}\right]^2-
2\alpha[1+\alpha\Upsilon(\vartheta)]\mathscr{I}(\vartheta,\theta) & = & 
\frac{1}{3}\frac{\Omega^2_{\rm o}R^3}{GM}\cos^3\theta[1+\alpha\Upsilon(\vartheta)] \\
\displaystyle \mathscr{I}(\vartheta,\theta) & = & \int_{\theta}^{\vartheta}\frac{\sin^{-1}x\cos^2x}{[1+\alpha\Upsilon(x)]^2}\left(\frac{d\Upsilon}{dx}\right)\,dx. 
\end{array} 
\label{solf_ta} 
\end{eqnarray}

\noindent As $\theta\to0$ and $\theta\to\pi/2$, also $\vartheta\to0$ and $\vartheta\to\pi/2$, respectively. Moreover, $\theta\to0$ both $\theta$ and $\vartheta$ are infinitesimals of the same order, so that we can write $\vartheta=p\theta$ where $p$ is constant. Expanding the function 
$\tan x$ for small values of $x$ and knowing that $\alpha\Upsilon<1$, we derive

\begin{eqnarray}
\begin{array}{rcl}
\displaystyle \ln\left(\frac{p\theta/2}{\theta/2}\right)+\cos(p\theta)-\cos\theta-\alpha[\Upsilon(p\theta)-\Upsilon(\theta)](1+...)-2\alpha[1+\Upsilon(p\theta)]\mathscr{I}(p\theta,\theta) & = &
\frac{1}{3}\eta\left(\frac{\Omega_{\rm o}}{\Omega_{\rm e}}\right)^2\!\!\!r^3\!\!\!\cos^3\theta[1+\alpha\Upsilon(p\theta)]^2.
\end{array} 
\label{lim1} 
\end{eqnarray}

\noindent Taking the limit of this expression for $\theta\to0$, and identifying $r(\theta=0$ with 
stellar surface it comes that

\begin{eqnarray}
\begin{array}{l}
\displaystyle \ln\frac{\vartheta}{\theta} = \frac{1}{3}\eta\left(\frac{R_{\rm p}}{R_{\rm e}}\right)^3
\left(\frac{\Omega_{\rm p}}{\Omega_{\rm c}}\right)^2 \\
\end{array}
\label{lim2} 
\end{eqnarray}

% \begin{eqnarray}
% \displaystyle \ln\frac{\vartheta}{\theta} = \frac{1}{3}\eta\left(\frac{R_{\rm p}}{R_{\rm e}}\right)^3\times
% \left[\begin{array}{ll}
% \displaystyle (1+\alpha)^2 &; \ \ {\rm for} \ \ \ \Upsilon(\theta)=\cos^k(\theta) \\
% \displaystyle (1+\alpha)^{-2} &; \ \ {\rm for} \ \ \ \Upsilon(\theta)=\sin^k(\theta) \\
% \end{array}
% \label{lim2} 
% \right]
% \end{eqnarray}

\noindent which introduced into Eq.~(\ref{f_th}) leads to

\begin{eqnarray}
\begin{array}{l}
\displaystyle \lim\limits_{\substack{\vartheta\to0 \\ \theta\to0}}f(r,\theta)  = f^{\rm pole} =
\exp\left]\frac{2}{3}\eta\left(\frac{R_{\rm p}}{R_{\rm e}}\right)^3\left(\frac{\Omega_{\rm p}}{\Omega_{\rm c}}\right)^2\right]
\end{array}
\label{tt1} 
\end{eqnarray}

% \begin{eqnarray}
% \begin{array}{l}
% \displaystyle \lim\limits_{\substack{\vartheta\to0 \\ \theta\to0}}f(r,\theta)  = f^{\rm pole} =
% \exp\left\{\frac{2}{3}\eta\left(\frac{R_{\rm p}}{R_{\rm e}}\right)^3\times\left[\begin{array}{ll}\displaystyle (1+\alpha)^2 & {\rm for} \ \Upsilon(\theta)=\cos^k(\theta) \\ 
% \displaystyle (1+\alpha)^{-2} & {\rm for} \ \Upsilon(\theta)=\sin^k(\theta) \\
% \end{array}
% \right]\right\}.
% \end{array}
% \label{tt1} 
% \end{eqnarray}

\noindent where $\Omega_{\rm p}$ is the polar angulat velocity, which carries that  $\Omega_{\rm p}/\Omega_{\rm e}=1+\alpha$ when $\Upsilon(\theta)=\cos^k(\theta)$ and $\Omega_{\rm p}/\Omega_{\rm e}=(1+\alpha)^{-1}$ for 
$\Upsilon(\theta)=\sin^k(\theta)$. We note that the dependence of $f^{\rm pole}$ with $\alpha$ and $k$ is given through the radii ratio $R_{\rm p}/R_{\rm e}$.\par
  To calculate the limit of $f(r,\theta)$ as $\theta\to\pi/2$ we make the change of variables 
$\vartheta\to\pi/2-\epsilon_{\vartheta}$ and $\theta\to\pi/2-\epsilon_{\theta}$, where
$\epsilon_{\vartheta}\to0$ and $\epsilon_{\theta}\to0$ are infinitesimals of the same order.
This operation implies the following changes of functions

\begin{eqnarray}
\begin{array}{lcl}
\displaystyle \tan\theta & = & \displaystyle \cot\epsilon_{\theta} \\
\displaystyle \cos\theta & = & \displaystyle \sin\epsilon_{\theta} = \epsilon_{\theta}-\epsilon^3_{\theta}/3!+...\\
\displaystyle \sin\theta & = & \displaystyle \cos\epsilon_{\theta} = 1-\epsilon^2_{\theta}/2+...\\
\displaystyle \ln\tan\frac{\theta}{2} & = & \displaystyle \ln(1-\tan\epsilon_{\theta}/2)-\ln(1+\tan\epsilon_{\theta}/2) \\
\displaystyle & = & \displaystyle -2\tan\frac{\epsilon_{\theta}}{2}-\frac{2}{3}\tan^3\frac{\epsilon_{\theta}}{2} \\
\displaystyle & = & \displaystyle -\epsilon_{\theta}-\frac{1}{6}\epsilon^3_{\theta}
\end{array}
\label{f_th2} 
\end{eqnarray}

\noindent and similar expressions for the functions dependent on $\vartheta$. Introducing this terms into Eq.~(\ref{solf_ta}), it is obtained

\begin{eqnarray}
\begin{array}{rcl}
\displaystyle \frac{1}{3}(\epsilon^3_{\theta}-\epsilon^3_{\vartheta})-(\epsilon_{\theta}-\frac{1}{6}\epsilon^3_{\theta}+...)[\alpha(\Upsilon_{\theta}-\Upsilon_{\vartheta})+...]-2\alpha(1+\alpha\Upsilon_{\vartheta})^2\int_{\pi/2-\epsilon_{\theta}}^{\pi/2-\epsilon_{\vartheta}}...d\,x & = &
\displaystyle \frac{1}{3}\eta\left(\frac{\Omega_{\rm o}}{\Omega_{\rm e}}\right)^2[1+\alpha\Upsilon(p\epsilon_{\theta})]^2
\end{array} 
\label{lim2} 
\end{eqnarray}

\noindent which for $\epsilon_{\theta}\to0$ reduces to

\begin{eqnarray}
\begin{array}{rcl}
\displaystyle \frac{1}{3}\left(1-\frac{\epsilon^3_{\vartheta}}{\epsilon^3_{\theta}}\right) & = &
\displaystyle \frac{1}{3}\eta\left(\frac{\Omega_{\rm o}}{\Omega_{\rm e}}\right)^2[1+\alpha
\lim_{\epsilon_{\theta}\to0}\Upsilon(p\epsilon_{\theta})]^2
\end{array} 
\label{lim3} 
\end{eqnarray}

\noindent where

\begin{eqnarray}
\begin{array}{rcl}
\displaystyle \lim_{\theta\to0}\Upsilon(p\epsilon_{\theta}) & = & \left\{\begin{array}{ll} 1 &; \displaystyle \ \ {\rm for} \ \Upsilon(\theta)=\cos^k(\theta) \\
\displaystyle 0 &; \ \ {\rm for} \ \Upsilon(\theta)=\sin^k(\theta) \\
\end{array}
\right.
\end{array} 
\label{lim4} 
\end{eqnarray}

\noindent and $\Omega_{\rm o}/\Omega_{\rm e}=1$ when $\Upsilon(\theta)=\cos^k(\theta)$, or
$\Omega_{\rm o}/\Omega_{\rm e}=1/(1+\alpha)$ if $\Upsilon(\theta)=\sin^k(\theta)$.\par
  Finally, with Eq.~(\ref{lim3}) and Eq.~(\ref{f_th}) we obtain 

\begin{eqnarray}
\begin{array}{lcl}
\displaystyle \lim\limits_{\substack{\vartheta\to\pi/2 \\ \theta\to\pi/2}}f(r,\theta) = 
f^{\rm eq} = (1-\eta{r^3})^{-2/3}  
\end{array}
\label{lim5} 
\end{eqnarray}

\noindent which is independent of $\alpha$ and $k$, and where for the stellar surface it is $\lim_{\theta\to\pi/2}r=1$ .\par
   The forms Eq.~(\ref{tt1}) and Eq.~(\ref{lim5}) can also be derived using the analytical expressions from Eq.~(\ref{tau_k1}) to Eq.~(\ref{tau_ks2}) by using series expansions for $\ln\tan{x}$, $\arctan{x}$ and $\ln(1\pm{x})$ valid for $x<1$ and relations in 
Eq.~(\ref{f_th2}).\par

\end{appendix}

\end{document}